\documentclass[prl,twocolumn,superscriptaddress]{revtex4-1}
\usepackage{amsmath}
\usepackage{amssymb}
\usepackage{bm}
\usepackage{epsfig}
\usepackage{graphicx}
\usepackage{color}
\usepackage[unicode=true,colorlinks=true,citecolor=blue,urlcolor=blue]{hyperref} 
\usepackage[utf8]{inputenc}
\usepackage[english]{babel}
\usepackage{xcolor}
\usepackage{gensymb}
\usepackage{physics}
\usepackage{ulem}
\usepackage{array}

\newcommand{\e}{\mathrm{e}}

\def\be{\begin{equation}}
\def\ee{\end{equation}}
\def\bea{\begin{eqnarray}}
\def\eea{\end{eqnarray}}
\def\beas{\begin{eqnarray*}}
\def\eeas{\end{eqnarray*}}
\DeclareMathOperator{\ctg}{\text{ctg}}

\begin{document}
\title{Strong enhancement of heavy-hole  Land\'e factor $q$ in InGaAs symmetric quantum dots revealed by coherent optical spectroscopy}
\author{A.~V.~Trifonov}
\email[correspondence address: ]{artur.trifonov@tu-dortmund.de}
\affiliation{Experimentelle Physik 2, Technische Universit\"at Dortmund, 44221 Dortmund, Germany}
\affiliation{Spin Optics Laboratory, St. Petersburg State University, 198504 St. Petersburg, Russia}
\author{I.~A.~Akimov}
\affiliation{Experimentelle Physik 2, Technische Universit\"at Dortmund, 44221 Dortmund, Germany}
\affiliation{Ioffe Institute, Russian Academy of Science, 194021 St. Petersburg, Russia}
\author{L.~E.~Golub}
\affiliation{Ioffe Institute, Russian Academy of Science, 194021 St. Petersburg, Russia}
\author{E.~L.~Ivchenko}
\affiliation{Ioffe Institute, Russian Academy of Science, 194021 St. Petersburg, Russia}
\author{I. A. Yugova}
\affiliation{Spin Optics Laboratory, St. Petersburg State University, 198504 St. Petersburg, Russia}
\author{A.~N.~Kosarev}
\affiliation{Experimentelle Physik 2, Technische Universit\"at Dortmund, 44221 Dortmund, Germany}
\affiliation{Ioffe Institute, Russian Academy of Science, 194021 St. Petersburg, Russia}
\author{S.~E.~Scholz}
\affiliation{Angewandte Festk\"orperphysik, Ruhr-Universit\"at Bochum, 44780 Bochum, Germany}
\author{C.~Sgroi}
\affiliation{Angewandte Festk\"orperphysik, Ruhr-Universit\"at Bochum, 44780 Bochum, Germany}
\author{A.~Ludwig}
\affiliation{Angewandte Festk\"orperphysik, Ruhr-Universit\"at Bochum, 44780 Bochum, Germany}
\author{A.~D.~Wieck}
\affiliation{Angewandte Festk\"orperphysik, Ruhr-Universit\"at Bochum, 44780 Bochum, Germany}
\author{D.~R.~Yakovlev}
\affiliation{Experimentelle Physik 2, Technische Universit\"at Dortmund, 44221 Dortmund, Germany}
\affiliation{Ioffe Institute, Russian Academy of Science, 194021 St. Petersburg, Russia}
\author{M.~Bayer}
\affiliation{Experimentelle Physik 2, Technische Universit\"at Dortmund, 44221 Dortmund, Germany}
\affiliation{Ioffe Institute, Russian Academy of Science, 194021 St. Petersburg, Russia}

\begin{abstract}

We reveal the existence of a large in-plane heavy-hole $g$ factor in symmetric self-assembled (001) (In,Ga)As/GaAs quantum dots due to warping of valence band states. This warping dominates over the well-established mechanism associated with a reduced symmetry of quantum dots and the corresponding mixing of heavy-hole and light-hole states. The effect of band warping is manifested in a unique angular dependence of a trion photon echo signal on the direction of external magnetic field with respect to the sample axes. It results in a uniform magnetic-field-induced optical anisotropy for the entire quantum dot ensemble which is a prerequisite for realization of spin quantum memories and spin-photon entanglement in the ensemble.

\end{abstract}

\date{\today}
\maketitle

In the field of quantum information, new applications based on spin qubits in solids are actively developed. Spin photonics studies based on coherent optical manipulation and measurement of spin qubits in semiconductor quantum dots (QDs) \cite{Imamoglu-review-NatPhot2015,Lodahl-NatNanoTech2018}, color centers in diamond \cite{Diamond} and SiC \cite{SiC} as well as rare-earth-ion doped crystals \cite{rare_earth} are heavily pursued. Here, the energy splitting of optical transitions into orthogonal linearly polarized spectral lines due to the Zeeman effect in a transverse magnetic field~\cite{Zeeman-Nature1897} is used to address the electron spin qubit and achieve spin-photon entanglement using properly polarized and frequency shaped optical fields~\cite{Imamoglu, Yamamoto, Steel}. To manipulate electron spins in a deterministic way, a precise knowledge of the energy splitting and magnetic-field-induced optical anisotropy, i.e. the orientation of eigen polarizations for optical transitions with respect to the direction of magnetic field, is required.

In atomic gases the energy splitting is proportional to the magnetic field strength, $B$, and the Land\'e $g$ factor, spectral lines are polarized either along ($\pi$) or perpendicular ($\sigma$) to the magnetic field axis~\cite{Zeeman-Nature1897}. In solids, the crystal field and localization potential lead to a modification of the $g$ factor requiring a description of the $g$ factor by a tensor. Consequently, the resulting axes of optical anisotropy (eigen polarizations) do not necessarily coincide with the magnetic field direction, but depend on the mutual orientation of the vector ${\bm B}$ and the sample axes~\cite{KusrayevPRL1999,Semenov-PRB2003, KoudinovPRB2004,Kiessling2006,Poltavtsev-PRR2020}. In direct band gap III-V and II-VI bulk semiconductors with zinc-blende lattice (as well as in group IV semiconductors with diamond lattice), the top valence band is formed by the heavy- and light-hole branches degenerate at zero wave vector ${\bm k} = 0$ ($\Gamma$ point) and the free-hole Zeeman splitting is dependent in a complicated way on the angle between the hole wave vector ${\bm k}$ and  magnetic field.

In low-dimensional systems with sizes on a nanometer scale, size-quantization results in a splitting of the bulk heavy- and light-hole branches into two series of hole subbands, $hh \nu$ and $lh \nu$ with $\nu= 1,2\dots$ In comparison with the T$_d$ point-group symmetry of bulk zinc-blende semiconductors, (001)-grown quantum wells have a reduced symmetry D$_{2d}$. As shown in Ref.~\cite{Marie1999}, the in-plane $g$ factor of a $hh1$ heavy hole is very small and given by the value of $3q$, where $q$ is one of the two bulk Land\'e factor parameters introduced by Luttinger \cite{Luttinger}. Most currently available self-assembled QDs  have a reduced symmetry C$_{2v}$ or lower than D$_{2d}$  \cite{KoudinovPRB2004,Kiessling2006,Nenashev-PRB2003,Zinovieva2003,Skolnick-PRB2005,Babinski,Belhadj-APL2010,Yakovlev2011,DomeShape2005,Glasman,Zunger,Kazimierczu-PRB2016,Belykh2016}.
There are several reasons for the symmetry reduction, in particular, the  asymmetry of a QD in the growth direction $z$ (e.g., pyramid-, lens- or dome-like shape of QDs), in-plane shape elongation, an in-plane strain etc., for more details see Ref.~\cite{Zunger}.
These strongly inhomogeneous factors produce a strong scatter in the polarization eigenstates in the QD ensemble and present the major obstacle for optical manipulation of a spin-qubit ensemble which requires uniform magnetic-field-induced optical anisotropy in all QDs.

In this letter we study self-assembled (In,Ga)As/GaAs QDs grown under special conditions. These QDs show the higher symmetry tetragonal point group D$_{2d}$ (or $\bar{4}2m$) that includes the mirror-rotation operation $S_4$ and, thereby, comprises symmetry along the growth axis. 
We have found that in these QDs, unlike in the (001)-grown quantum wells, the in-plane hole $g$ factor exceeds by far the bulk GaAs value $|3q|$. To explain this unexpected finding we propose a new mechanism contributing to the enhancement of heavy-hole Land\'e factor parameter $q$. The enhancment stems from the strong localization of the hole within a QD and is governed by the difference $\gamma_3 - \gamma_2$ of Luttinger valence band parameters~\cite{Luttinger} which is responsible for the bulk valence band warping. The new mechanism is shown to dominate in symmetric QDs and leads to a uniform magnetic-field-induced optical anisotropy in the entire QDs ensemble. For the experimental confirmation, we have studied the coherent optical QD response in the form of spin dependent photon echoes from trions~\cite{Poltavtsev-PRR2020} in singly electron charged QDs. The high symmetry of the QDs is confirmed by the dependence of photon echo signal on the orientation of external magnetic field with respect to the sample axes. The obtained in-plane hole $g$ factor value of about 0.2 associated with the proposed mechanism is comparable with that of the conduction band electron $g$ factor.

We study singly electron charged QDs and analyze the spin properties of a resident electron and a hole (in the trion) occupying the QD ground states $e1$ and $hh1$, respectively. 
First we perform a symmetry analysis of the Zeeman Hamiltonian for a nanostructure of the point-group D$_{2d}$. Then we analyze the consequences of possible symmetry-breaking distortions. In a structure of D$_{2d}$ symmetry, the $e1$ conduction-electron and $hh1$ heavy-hole states transform according to the equivalent representations $\Gamma_6$ as the spinors $\psi^e_{1/2} =\uparrow \! S$, $\psi^e_{-1/2} = \downarrow \! S$ and pair of functions $\psi^h_{1/2} = \downarrow \! (X - {\rm i} Y)/\sqrt{2}$, $\psi^h_{-1/2} = -\uparrow \! (X + {\rm i} Y)/\sqrt{2}$, see \cite{Glasman,IvchPikusBook}. Here $S$ and $X,Y$ are, respectively, the conduction-band and valence-band Bloch functions at the $\Gamma$ point. In the chosen bases, the  Zeeman Hamiltonian matrices in the magnetic field ${\bm B} \perp z$ have the same structure
\begin{eqnarray} \label{HeHh}
&&{\cal H}^e ({\bm B})= \frac12\: \mu_B g_e (\sigma_x B_x + \sigma_y B_y),\\ &&
{\cal H}^h ({\bm B})= \frac12\: \mu_B g_h (\sigma_x B_x + \sigma_y B_y), \nonumber
\end{eqnarray}
and differ only in the values of the in-plane $g$ factors, $g_e$ and $g_h$. Hereafter $\mu_B$ is the Bohr magneton, $x \parallel [100], y \parallel [010]$, and $\sigma_x, \sigma_y$ are the Pauli 2$\times$2 matrices which coincide for electron and heavy hole. Note that, in the other frequently used hole basis $\tilde{\psi}^h_{1/2}=-\uparrow(X + {\rm i} Y)/\sqrt{2}$, $\tilde{\psi}^h_{-1/2}=\downarrow(X - {\rm i} Y)/\sqrt{2}$, the second term in ${\cal H}^h ({\bm B})$ has the opposite sign. 

\begin{figure}
\includegraphics[width=\linewidth]{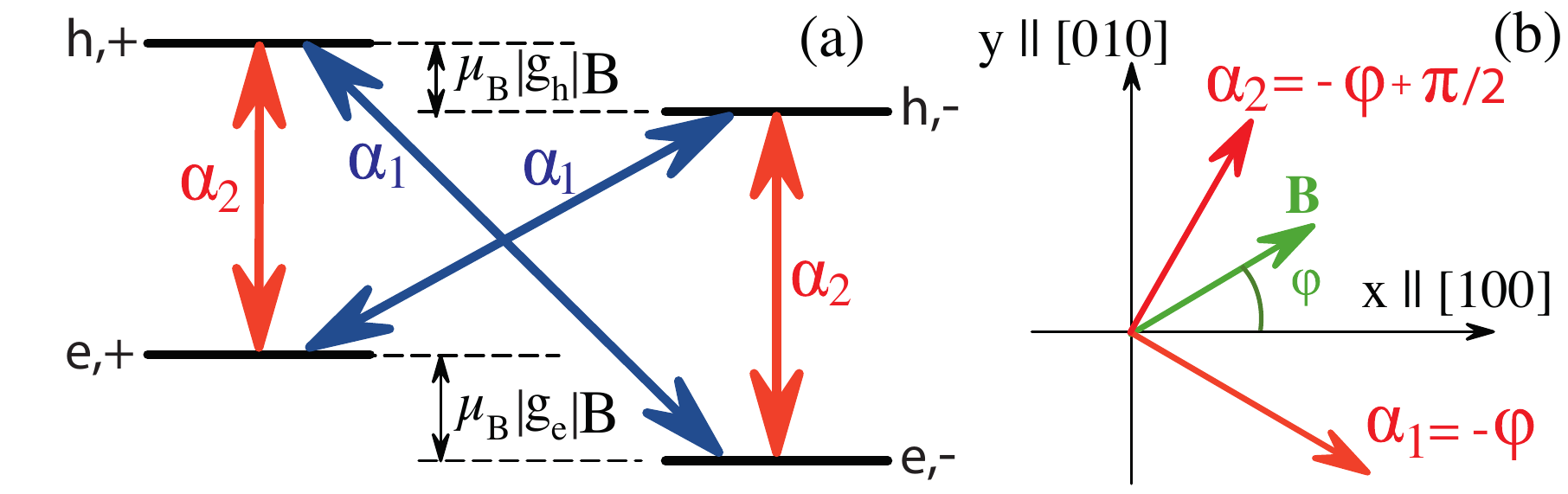}
\caption{
(a) Schematics of sublevels of resident electron and trion (hole and two singlet electrons) split in the in-plane magnetic field. Optical transitions indicated by the arrows are linearly polarized with directions given by angles $\alpha_1$ and $\alpha_2$ in Eq.~(\ref{gamma12}) for $g_e g_h>0$. If $g_e g_h<0$ then $\alpha_1$ and  $\alpha_2$ are exchanged. (b) Schematics of direction of external magnetic field ${\bm B}$ and directions of eigen polarizations $\alpha_1$ and $\alpha_2$.
}
\label{Fig1}
\end{figure}

The magnetic field splits the electron and hole spin states into the energy sublevels $E^i_{\pm} = \pm \mu_B |g_i| B/2$ ($i = e, h$). The selection rules for the optical transitions from the electron sublevel $E^e_{\pm}$ to the trion state with a pair of  singlet electrons and a hole in the sublevel $E^h_{\pm}$ are shown in Fig.~\ref{Fig1}(a) for $g_e g_h>0$. The optical transitions are linearly polarized along the directions determined by the angles $\alpha_{1,2}$ between the polarization unit vector ${\bm e}$ and the $x$ axis. These angles are related to the angle $\varphi$  between the magnetic field vector and the $x$ axis (Fig.~\ref{Fig1}(b))  by
\begin{equation} \label{gamma12}
\alpha_1(D_{2d}) = - \varphi, \quad \alpha_2(D_{2d}) = - \varphi + \frac{\pi}{2},
\end{equation}
where D$_{2d}$ indicates the QD symmetry. In spite of the isotropic Hamiltonians (\ref{HeHh}), the behaviour of $\alpha_1$ or $\alpha_2$ as a function of $\varphi$ reveals the tetragonal symmetry. Particularly,  
for $g_e g_h > 0$ the transition $(e,+) \to (h,+)$ is polarized perpendicular to the magnetic field if ${\bm B} \parallel [100], [010], [\bar{1}00]$ or $[0\bar{1}0]$,  while, for ${\bm B} \parallel [110], [\bar{1}10], [\bar{1}\bar{1}0]$ or $[1\bar{1}0]$, it is polarized along ${\bm B}$. The D$_{2d}$ symmetry Hamiltonians (\ref{HeHh}) leads to a variation of the optical polarization described by the fourth harmonic as function of $\varphi$ \cite{Semenov-PRB2003}.

If a nanostructure is distorted by a perturbation of the symmetry $B_1$ (like $x^2 - y^2$) and/or $B_2$ (like $2xy$), where $B_1$ and $B_2$ are the irreducible representations of the D$_{2d}$ group, then the Zeeman Hamiltonians have additional anisotropic contributions
\begin{equation} \label{HeHh2}
{\cal H}_{\rm an}^i ({\bm B})= \frac{\mu_B}{2} [g_{1i} (\sigma_x B_x - \sigma_y B_y) + g_{2i} (\sigma_x B_y + \sigma_y B_x)],
\end{equation}
where $g_{1i}$ and $g_{2i}$ relate to $B_1$ and $B_2$.
The spin-split states have the energies $E^{e,h}_{\pm} = \pm \hbar \omega_i /2$ with the spin splitting given by $\hbar \omega_i  =  \tilde{g}_i \mu_B B$, where
\begin{equation} \label{energy}
\tilde{g}_i = \sqrt{g_i^2 + g_i^{\prime 2}+ 2 g_i g'_i \cos{[2 (\varphi} - \chi_i)]} ,
\end{equation}
$g_i^{\prime} = \sqrt{g_{1i}^2 + g_{2i}^2}$ and $2 \chi_i = \arctan{(g_{2i}/g_{1i})}$.

Because of the anisotropy caused by the distortion (\ref{HeHh2}), the effective magnetic field $\tilde{\bm B}$ acting on the carrier is directed not along the vector ${\bm B}$ and has the angle
\begin{equation} \label{thetai}
\theta_i = \arg{\{g_i  {\rm e}^{{\rm i} \varphi} + (g_{1i} + {\rm i}  g_{2i}) {\rm e}^{-{\rm i} \varphi}\}}
\end{equation}
with the $x$ axis, the values of $\theta_i$ can cover the full circle (0, $2 \pi$). The spin-split eigen states are given by 
\begin{equation}
\ket{\psi^i_{\pm}} = \frac{1}{\sqrt{2}}\left( \mathrm{e}^{ - \mathrm{i} \theta^i_{\pm} /2} \psi^i_{1/2} + \mathrm{e}^{ \mathrm{i} \theta^i_{\pm}/2 } \psi^i_{-1/2} \right) \varphi_i({\bm r}),
\end{equation}
where $\varphi_e({\bm r})$ and $\varphi_h({\bm r})$ are the $e1$ and $hh1$ envelope functions, $\theta_+^i = \theta_i$ and $\theta_-^i = \theta_i+\pi$. 

The optical transitions $e,\pm \to h, \pm$ and $e,\pm \to h, \mp$ are also linearly polarized as for a QD of D$_{2d}$ symmetry. However, the corresponding eigen polarizations are now determined not by Eq.~(\ref{gamma12}) but by the more general equations
\begin{equation}
\label{alpha1}
\alpha_1 = - \frac{\theta_e + \theta_h}{2}, \quad \alpha_2 = - \frac{\theta_e + \theta_h - \pi}{2}.
\end{equation}

For the experimental study of the Zeeman effect in transverse magnetic field we use an approach based on spin dependent photon echoes (PE)~\cite{Poltavtsev-PRR2020}. The advantage of this technique is the unique possibility of obtaining the full set of Zeeman splittings and optical anisotropy even if they are hidden by inhomogeneous broadening of the optical transitions. We study self-assembled (In,Ga)As/GaAs QDs grown by molecular beam epitaxy with subsequent annealing procedure as described in SM section I~\cite{SupplMat}. In order to increase light-matter coupling and PE signal amplitude~\cite{Poltavtsev-PRB2016,Salewski-PRB2017,Kasprzak-Optica2018} four QD layers are placed in the antinodes of a standing electromagnetic wave of a weak-coupling microcavity with the quality factor $Q\sim1000$~\cite{Kamenskii-PRB2020}. Modulation doping with Si provides one resident electron to each QD on average.

\begin{figure}
\includegraphics[width=\linewidth]{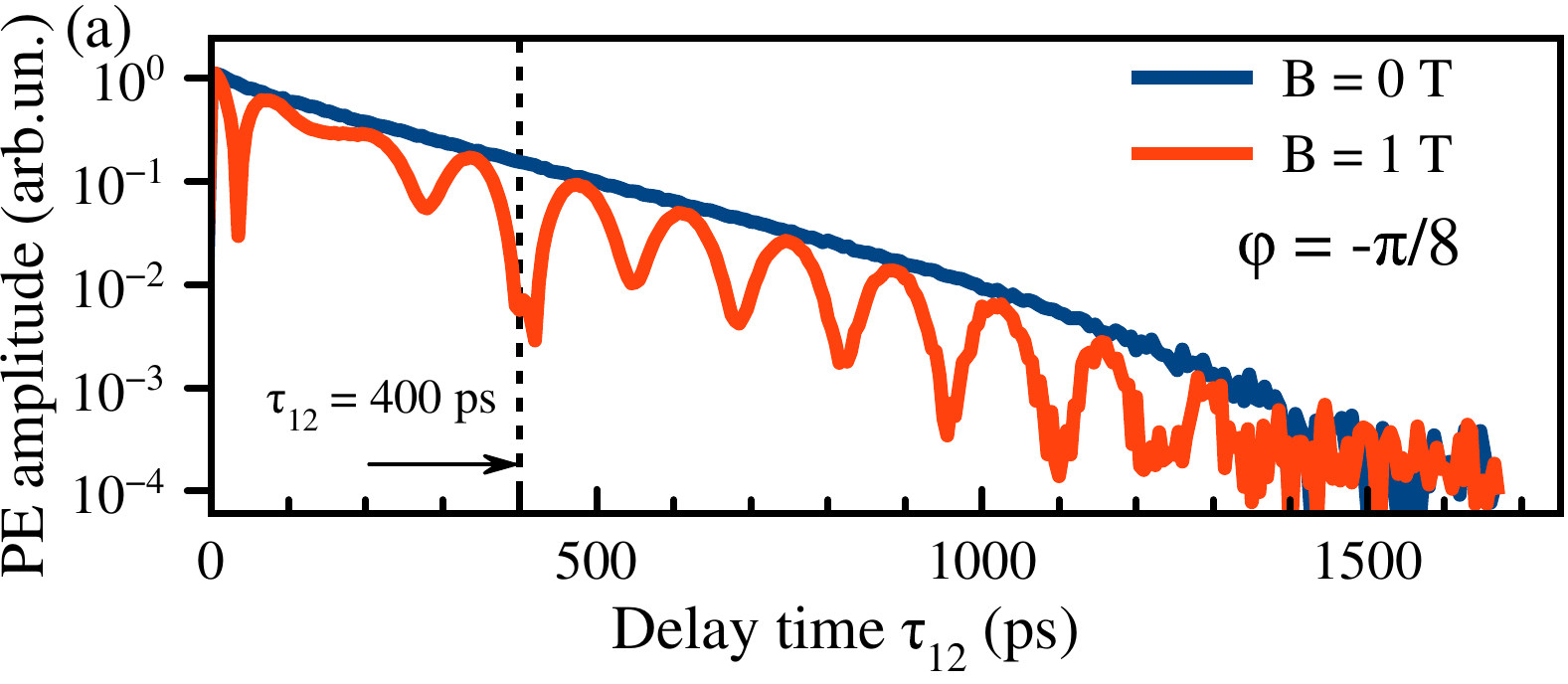}
\includegraphics[width=\linewidth]{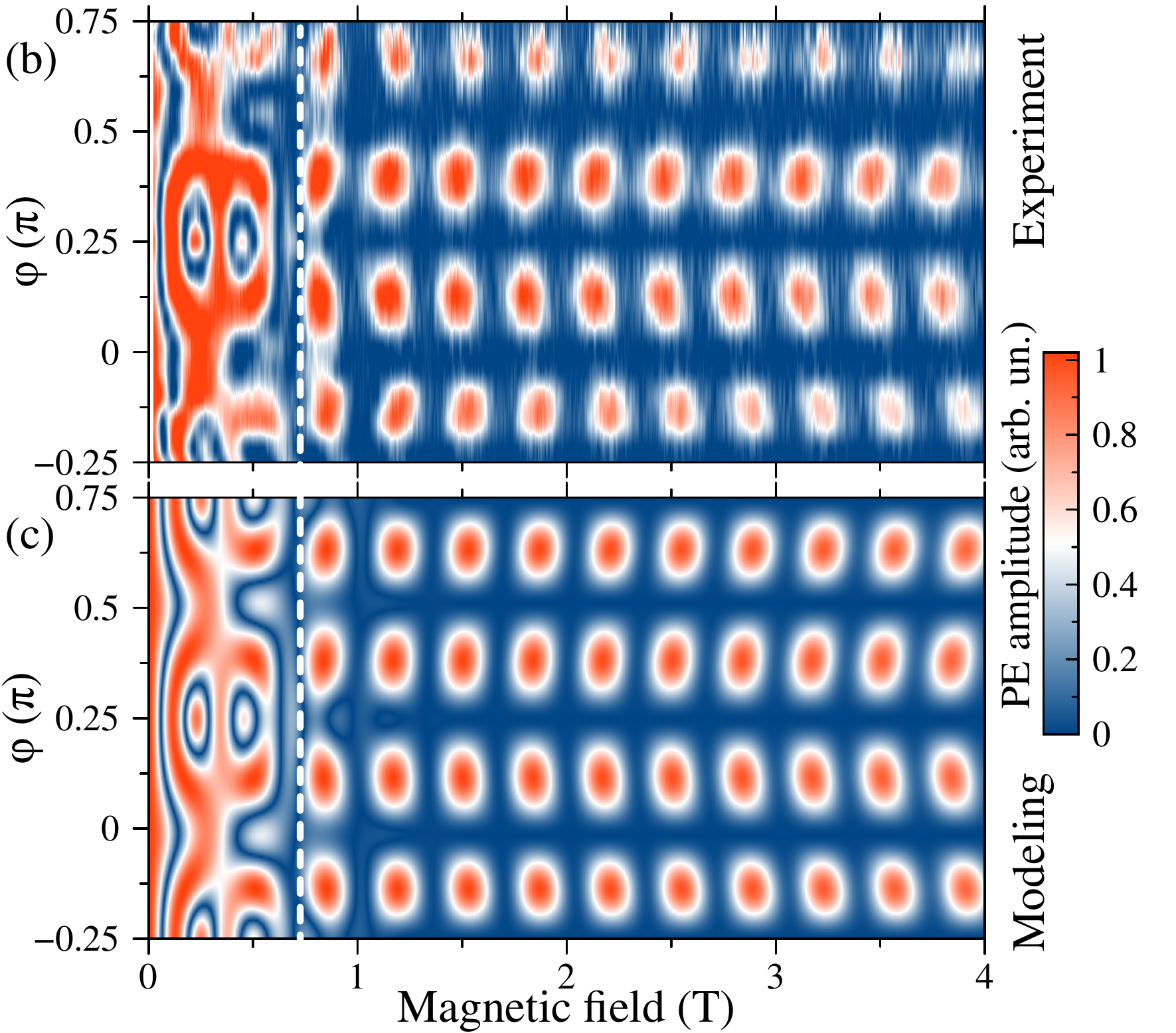}
\caption{(a) Photon echo amplitude of (In,Ga)As/GaAs QDs as a function of $\tau_{12}$. The transients are measured for $\varphi = -\pi/8$ at $B=0$ and 1~T.  (b) Photon echo amplitude measured as a function of $B$ and $\varphi$ at $\tau_{12}= 400$~ps in the HVH  polarization configuration, T=1.4~K. (c) Corresponding calculations using $T_2 = 430$~ps, $g$ factors from Fig.~\ref{FigDecay}, their spread $\Delta g_e= 0.005$, $\Delta g_h = 0.065$ for electrons and holes, respectively. Detailed evaluation of the parameters used in calculations are presented in SM section IV~\cite{SupplMat}.
Note that we measure the absolute value of PE amplitude and therefore the calculations show $|P_{HVH}|$.
}
\label{FigColorPlot}
\end{figure}

The sample is placed into a superconducting split-coil cryostat and kept at the temperature of 1.4~K. The magnetic field is applied in Voigt geometry in the $xy$-plane and rotation of the sample around $z$-axis allows us to vary the angle $\varphi$. A sequence of two optical pulses with $2$ ps duration delayed by the $\tau_{12}$ time with respect to each other excites the QDs under nearly normal incidence (see details in SM section II~\cite{SupplMat})). The photon energy is tuned into resonance with the cavity mode and set to 1.434~eV. The transient four-wave mixing signal is detected in reflection geometry using heterodyne detection~\cite{PSS2018}. Due to inhomogeneous broadening of the optical transitions in the ensemble of QDs the signal is represented by a photon echo  which is delayed by $2\tau_{12}$ with respect to the first excitation pulse~\cite{Poltavtsev-PRB2016}. 
At $B=0$ the PE amplitude decays exponentially $\exp(-2\tau_{12}/T_2)$ with the optical coherence time $T_2=430$~ps as shown in Fig.~\ref{FigColorPlot}(a). For $B\ne0$ the PE signal shows oscillations due to spin precession of electrons and holes. 
Such spin dependent PE signal is sensitive to the polarization configuration of the excitation pulses~\cite{LangerPRL}. Using linearly polarized optical pulses allows us to determine precisely the eigen polarizations $\alpha_{1,2}$ as a function of $\varphi$~\cite{Poltavtsev-PRR2020}. In what follows we concentrate on the HVH polarization configuration where the first pulse is polarized along the horizontally (H) oriented field ${\bm B}$, while the polarization of the second pulse is vertical (V). The detection of PE is performed in H polarization.

Figure~\ref{FigColorPlot}(b) shows experimental data for the spin dependent PE amplitude as a function of $\varphi$ and $B$ at a fixed value $\tau_{12} = 400$~ps  (vertical dashed line in Fig.~\ref{FigColorPlot}(a)) measured in a steps of $\pi/18$. Two types of oscillations are observed. First, there are oscillations along the $B$-axis due to variation of Larmor precession frequencies of electrons $\omega_e$ and holes $\omega_h$.  Second, angular $\varphi$ oscillations appear because of the dependence of $\alpha_{1,2}$ on $\varphi$. The signal behaves differently for the ranges $B\le0.7$~T and $B\ge0.7$~T separated by the vertical dashed line in Fig.~\ref{FigColorPlot}(b). 
This is attributed to the large spread of hole $g$ factor, $\Delta g_h$, which results in decay of the hole spin precession contribution to the PE signal. Nevertheless, optical anisotropy can be evaluated from the angular dependence even for large $B$, where the PE amplitude $P_{HVH}$ is decribed by the simple relation (see SM section III~\cite{SupplMat})
\begin{equation}
P_{HVH} \sim \left (1-\cos{[4(\alpha_1 - \varphi)]} \right) \sin^2{(\omega_e \tau_{12}/2)}.
\label{PHVHinh}
\end{equation}
In this case, the magnetic field and the delay $\tau_{12}$ oscillations of the PE signal are associated only with the electron spin precession. It follows from Eq.~({\ref{PHVHinh}}), that the D$_{2d}$ symmetry contribution ($\alpha_1 = -\varphi$) gives rise to the eighth harmonic in the $P_{HVH}(\varphi)$ dependence. By contrast, for  the C$_{2v}$ symmetry the hole contribution $\alpha_1 = -\chi_h/2$ (see Eqs.~(\ref{energy}),(\ref{alpha1})) the angular dependence contains the fourth harmonics.  

The angular dependence in Fig.~\ref{FigColorPlot}(b) at $B>0.7$~T shows four oscillations within the range  $0 \le \varphi \le \pi$, i.e. we observe the eighth harmonics. The contrast of oscillations $C= (P_{\rm max}-P_{\rm min})/(P_{\rm max}+P_{\rm min}) \approx 0.95$ is very high, where $P_{\rm max}$ and $P_{\rm min}$ are the maximum and minimum values of $|P_{HVH}|$. Thus we conclude  that the D$_{2d}$ symmetry gives the main contribution to the hole $g$ factor. Moreover, the high contrast of angular oscillations indicates that the spread of directions of eigen-polarizations (spread of $\alpha_1$) in the QDs ensemble under study is small~\cite{SupplMat}. 

\begin{figure}
\includegraphics[width=0.9\linewidth]{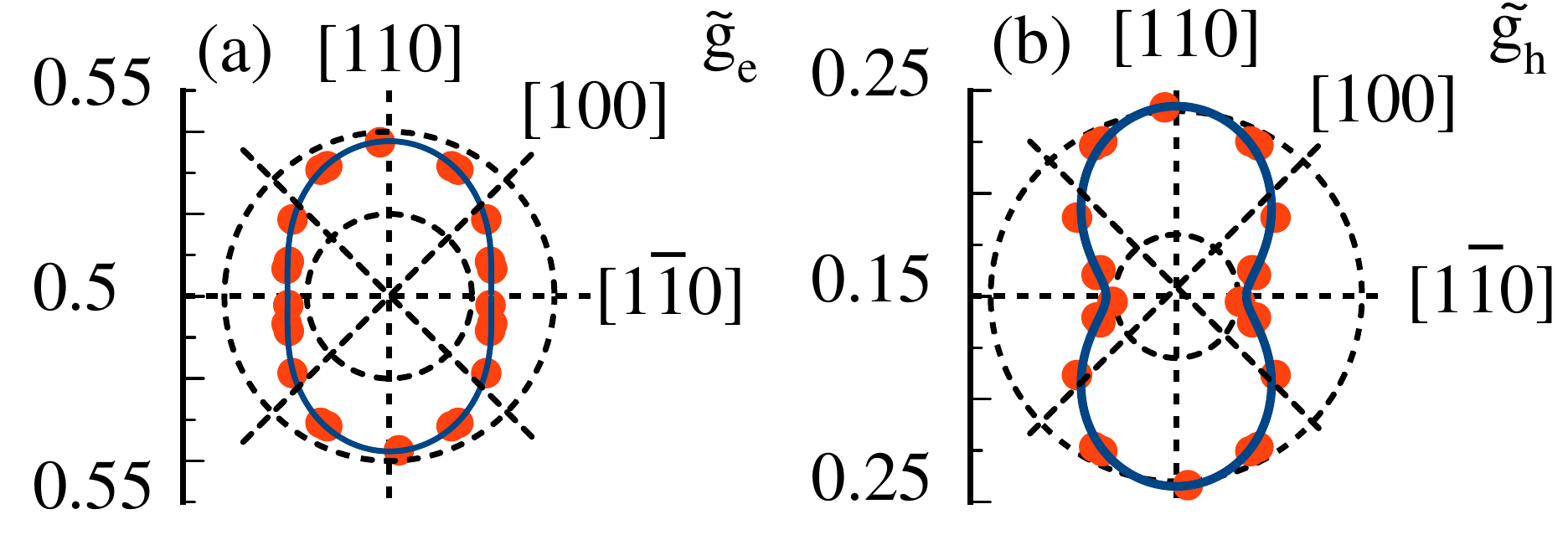}
\caption{Angular dependence of the (a) electron $g$ factor $\tilde{g}_e$ and (b) the hole $g$ factor $\tilde{g}_h$ obtained from fitting of the experimental data (dots) with Eq.~(\ref{energy}) (solid lines). Dashed circles describe minimum and maximum electron $g$ factor values, 0.52 and 0.54, in the panel (a) and hole $g$ factor values, 0.18 and 0.24, in the panel (b). Dashed straight lines indicate orientations of magnetic field with respect to the crystallographic axes. 
}
\label{FigDecay}
\end{figure}

In order to determine accurately the dependences $\tilde{g}_{i} = \hbar \omega_{i}/\mu_B B$ on $\varphi$ we analyze the PE transients for different values of $B$, polarization configurations and sample orientations. The details of fitting are presented in SM section IV~\cite{SupplMat}. The analysis shows no dependence of $\tilde{g}_{e,h}$ on magnetic field strength. The obtained angular dependences of electron and hole $g$ factors are shown in Fig.~\ref{FigDecay} by red dots. 
 As one can see, the value of $\tilde{g}_e$ changes between $0.52$ and $0.54$ and can be fitted by Eq.~(\ref{energy}) (solid line) with parameters $g_e = -0.531$, $g'_e =0.007$, $\chi_e = 0$, where we take into account that in (In,Ga)As QDs ${g}_e<0$~\cite{Belykh2016,Nakaoka-PRB2005,YugovaPRB2005}.  Thus, the direction of the effective magnetic field ${\tilde{\bm B}}$ negligibly deviates from ${\bm B}$ by less than $1^\circ$. 

The angular dependence of $\tilde{g}_h$ clearly indicates that both D$_{2d}$ and low-symmetry  contributions have impact on the hole's $g$ factor~\cite{Semenov-PRB2003, Poltavtsev-PRR2020}. Here, a weak low-symmetry contribution is added to the dominant D$_{2d}$ contribution. The dependence of $\tilde{g}_h(\varphi)$ can be approximated by Eq.~(\ref{energy}) with $g_h = 0.213$, $g'_h = 0.029$, and $\chi_h = \pi/2$, as shown in Fig.~\ref{FigDecay}(b). The positive sign of  $g_h$ follows from the theoretical model which is presented below. The calculated dependencies of PE amplitude on $B$ and $\varphi$ with the obtained electron and hole $g$ factor tensors are shown in Fig.~\ref{FigColorPlot}(c). The excellent correspondence between experimental  and  theoretical figures in Figs.~\ref{FigColorPlot}(b) and \ref{FigColorPlot}(c) confirms the accuracy of the analysis above.

The obtained values of $g_h$ and $g'_h$ are non-trivial. In a zinc-blende quantum well (QW) structure grown along the [001] direction $g_h=3 q$,  where $q$ is the negative Land\'e valence-band parameter of the bulk semiconductor (in Ref.~\cite{Marie1999} a value of $-3q$ is used with positive $q$). One of the aims of our work is to demonstrate that in a quantum dot having D$_{2d}$ symmetry the quantum confinement in the $xy$-plane can result in a remarkable enhancement of the factor $|3q|$.  This can be understood taking into account the expansion of
the heavy-hole Zeeman Hamiltonian in powers of wave vector ${\bm k}$  \cite{Marie1999}
\begin{equation} \label{expansionH}
{\cal H}({\bm B},k^2) = \frac12 \mu_B (3 q + ck^2 + \dots) (\sigma_xB_x + \sigma_y B_y),
\end{equation}
resulting in the renormalization of the hole $g$ factor given by  $g_h({\bm k}) = 3 q + c k^2 + \dots$
The coefficient $c$ can be conveniently presented as $G (\gamma_3 - \gamma_2) L_W^2$, where 
$\gamma_2$ and $\gamma_3$ are the dimensionless Luttinger valence-band parameters and $L_W$ is the QW width, for a GaAs-based QW the coefficient $G$ takes the value of 0.08 (see details in SM section V~\cite{SupplMat}). The factor ${\gamma_3 - \gamma_2}$ shows that $c$ is related to the bulk valence band warping. At low temperature the value of $c k^2$ in a QW is small compared to $3q$ and can be ignored~\cite{Marie1999}. In a QD, $k^2$ should be averaged over the quantum-confined state leading to 
\begin{equation} \label{expansiong}
g_h = 3(q + q_w),
\end{equation}
where $3 q_w = c \left< k^2 \right>$, and therefore the parameter $q$ is renormalized by a factor of $1 + (q_w/q)$. For a quantum dot-disk of radius $R$ we obtain
$3 q_w \approx 0.4\ (L_W/R)^2$. For a parabolic GaAs-based  QD with the confining potential $V({\bm r}) = [\kappa_z z^2 + \kappa_{\parallel}(x^2 + y^2)]/2$, we have $3 q_w \approx 0.5 \sqrt{\kappa_{\parallel}/\kappa_z}$.
For the aspect ratios $2R$:$L_W$=3:1 and $\kappa_z$:$\kappa_{\parallel}$= 3:1, the quantum-confinement contribution $3 q_w \approx 0.3$ by far exceeds the experimentally measured bulk value $|3 q| \approx 0.035$~\cite{Marie1999}. Note that, with decreasing $R$ values or increasing $\kappa_{\parallel}$ values
 the higher orders terms in the expansion~(\ref{expansionH})  should be also taken into account and the above estimates of $q_w$ give only its order of magnitude.

In the existing theories of the heavy-hole in-plane Land\'e factor in QWs and QDs of the  symmetry C$_{2v}$ or lower, the values  $g_{1h}$ or $g_{2h}$ are determined by the heavy-light hole mixing induced by the distortions \cite{Semenov-PRB2003,Nenashev-PRB2003,Zinovieva2003,KoudinovPRB2004,Kazimierczu-PRB2016,PikusPikus}. In the proposed enhancement of the parameter $q$ in QDs of D$_{2d}$ symmetry, the Bloch heavy- and light-hole functions are naturally mixed by the hole nonzero wave vectors ${\bm k}$ and the quantization of $k^2$ in QDs causes the renormalization of $q$.

The theory gives two important predictions. First, because of opposite signs of $q$ and $c$, there are QDs with larger base size where $3q$ and $3q_w$ compensate each other and the in-plane $g$ factor vanishes. Second, besides the term $c k^2$ in Eq.~(\ref{expansionH}), there is an additional term $\delta {\cal H} ({\bm B},{\bm k}) =(\mu_B/2)c' (\sigma_+ B_+ k_-^2 + \sigma_- B_- k^2_+)$ \cite{Marie1999,Miserev}, where $\sigma_{\pm} = (\sigma_x \pm {\rm i} \sigma_y)/2$ and $c' = [(\gamma_3 + \gamma_2)/(\gamma_3 - \gamma_2)]c$. In a QD of D$_{2d}$ symmetry this term does not contribute to the hole $g$ factor. However, for a QD shape of reduced symmetry the average values of $\langle k^2_x - k^2_y \rangle$ and $\langle 2 k_x k_y \rangle$ do not vanish and the corresponding terms have an impact to low symmetry contribution through the coefficients $g_{1h}$ and $g_{2h}$ in Eq.~(\ref{HeHh2}).

In conclusion, we have revealed experimentally and theoretically that the surprisingly large in-plane hole $g$ factor in an ensemble of strongly annealed (In,Ga)As/GaAs QDs is dominated by the D$_{2d}$ symmetry contribution. The proposed enhancement of the Land\'e valence-band parameter $q$ in QDs of the D$_{2d}$ symmetry allows us to explain the unique angular patterns of spin dependent photon echoes in the in-plane magnetic field. The enhancement appears because of the in-plane confinement of holes and the valence band warping. It results in the uniform magnetic-field-induced optical anisotropy for the entire quantum dot ensemble which is highly appealing for application in quantum information devices.

\begin{acknowledgments}
The authors acknowledge financial support by the Deutsche Forschungsgemeinschaft through the International Collaborative Research Centre TRR 160 (Projects A3 and A1). A.V.T. and I.A.Y. thank the Russian Foundation for Basic Research (Project No. 19-52-12046) and the Saint Petersburg State University (Grant No. 73031758). A.L. and A.D.W. gratefully acknowledge financial support from the grants DFH/UFA CDFA05-06, DFG project 383065199, and BMBF Q.Link.X 16KIS0867.
L.E.G and E.L.I. thank the Russian Foundation for Basic Research (Project No. 19-52-12038). L.E.G. was supported by the Foundation for the Advancement of Theoretical Physics and Mathematics ``BASIS''.
\end{acknowledgments}

\pagebreak

\begin{widetext}

\section{Supplementary materials}

\section{Sample}

The sample under study (No. 14833 RTA 900$^\circ$C) was grown by the molecular beam epitaxy method on [100] GaAs substrate. The sample contains a $5/2 \lambda$ microcavity formed by 14 GaAs/AlAs pairs in the bottom distributed Bragg reflector (DBR), and 11 pairs in the top DBR. These parameters correspond to a theoretical Q factor $\sim1000$. Four layers of (In,Ga)As/GaAs quantum dots are placed in all four antinodes of the microcavity standing electromagnetic wave. The $\delta$ layer of Silicon doping is located at 64.3 nm under each QDs layer (half the distance between the QDs layers). 

The QDs were grown at a pyrometer temperature reading of 515~$^\circ$C to 520~$^\circ$C by depositing an amount of nominally 1.7 mono layers of InAs at a deposition rate of 0.005 nm/s. After a 20 s annealing break and another break of 35 s at a pyrometer reading of 495~$^\circ$C, the QDs were overgrown by GaAs at the latter temperature. 
This low temperature maintains partially the shape of the QDs during overgrowth. 
Typically, anisotropic surface diffusion leads to a slight shape anisotropy. After the sample is grown and characterized by photoluminescence, rapid thermal annealing at 900~$^\circ$C for 30 s of small sample pieces (5x5 mm) was performed. During this step, Indium diffuses out of the QDs and Gallium diffuses in. This results in a larger band gap material and thus a strong blue shift. The annealing parameters were adapted such that the blueshift matches the microcavity photon mode.
The Indium-Gallium interdiffusion during annealing is in part driven by strain and tends to homogenize the Indium content. From a half-lens shaped form of the QDs before annealing, an oblate spherical form develops. The inhomogeneous broadening of the ensemble reduces, as smaller dots (originally on the high energy side of the spectrum) are less strained and less diffusion takes place, while larger dots strongly drive diffusion. Shape imperfections like assymmetries are reduced during the annealing procedure, as the diffusion blurs the shape to the aforementioned oblate sphere. Moreover, hard boundaries as present in as grown or Indium-flush QDs are smoothed out by this blurring as well.

Because of a gradient of growth parameters on the sample area, there is a dependence of optical resonance energies of  the QDs as well as the photonic mode of the microcavity on the sample point. For the experimental study, we chose a sample point with microcavity resonance at 1.434 eV, at which the PE signal was maximal.  Details of optical characterisation of the sample can be found in Ref. [33].

\section{Experimental details}

The sample was cooled down in a liquid helium bath cryostat to a temperature of 1.4~K. The cryostat was equipped with a superconducting magnet, allowing one to perform the optical experiments in the Voigt geometry (magnetic field is almost orthogonal to the optical axis). A piezo-mechanical translator (attocube) allows rotating the sample around the axis perpendicular to the sample plane (rotating around the optical axis). All the optical pulses are emitted by a picosecond Ti:Sapphire laser which generates pulses of $\sim 2$~ps duration with a repetition rate of 75.75~MHz. The time delays between the pulses are changed using mechanical delay lines.

The optical pulses were focused to a spot of about 200~$\mu$m diameter using an 0.5~m spherical metallic mirror. The incidence angles of pulses are close to normal and equal to $\approx 1/50$~rad and $\approx 2/50$~rad (correspond to in-plane wavevectors $k_1$ and $k_2$). The PE pulses were measured in reflectance geometry in direction $\approx 3/50$~rad which corresponds to the PE wavevector $2k_2-k_1$. Optical heterodyne detection was used to perform time-resolved PE experiments and to enhance detected signals. By mixing with a relatively strong reference pulse and scanning the time delay between the first pulse and the reference pulse, $\tau_{ref}$, one can measure the temporal profile of the photon echo pulse. The simultaneous scan of $\tau_{12}$ and $\tau_{ref}=2\tau_{12}$ allows one to measure the decay and spin dynamics of the PE as shown in Fig.~2(a) of the main text.

The typical polarizations of the pulses in photon echo experiments are horizontal (H) and vertical (V) where the external magnetic field is directed horizontally parallel to the H polarization. Note, that the reference pulse polarization determines the polarization of photon echo detection. Here we use a three-letter notation for the polarization configuration of the photon echo experiment. HHH, for example, indicates that the polarizations of all three pulses (two pulses hitting the sample and the polarization of detection) are horizontal. In the HVH configuration, the second pulse has vertical polarization, while the first and reference pulses have horizontal polarization. 

The experimental study of anisotropy by the photon echo method was performed by rotating the sample, which is equivalent to changing the angle $\varphi$ by rotation of ${\bm B}$. In experiment the crystallographic axis was determined with respect to cut edge of the sample which has to be close to the crystallographic axis [110].

\section{Modelling of spin dependent photon echo}

The modelling of photon echo signals is based on the Liouville equation:
\begin{equation}
\mathrm{i} \hbar \dot{\rho} = \left[H,\rho \right] + \Gamma,
\end{equation}
where $\rho$ is the density matrix describing the system state, $H$ is the Hamiltonian, $\Gamma$ describe the decay phenomenologically. The Hamiltonian consists of three terms $H = H_0+ H_B+V$, where $H_0$ describes the system without magnetic field and interaction with light, $H_B$ corresponds to influence of magnetic field, and $V$ takes into account interaction with light. Details of the problem, the form of operators, as well as solution of the problem can be found in Ref.~[14]. Here we introduce the final equations which describe the detected in experiment signals of two-pulse spin dependent photon echo.  It is worth noting that in Ref. [14] for hole state the basis $-\uparrow$$(X + {\rm i} Y)/\sqrt{2}$, $\downarrow$$(X - {\rm i} Y)/\sqrt{2}$ was used. Therefore $\varphi_h$ from Ref.~[14] is related with $\theta_h$  by $\varphi_h=-\theta_h$. Thus, the equation for the case of linearly polarized optical pulses reads
\bea
\e^{\frac{2\tau_{12}}{T_2}}P &\sim& \left [ \sin{(\omega^0_e \tau_{12})}\sin{(\omega^0_h \tau_{12})} \cos{(2 p_2-2\alpha_0)}
+1-\cos{(\omega^0_e \tau_{12})} \cos{(\omega^0_h \tau_{12})}  \right ] \cos{(p_3-p_1)}  \nonumber \\
&+& \left [ (1+\cos{(\omega^0_e \tau_{12})} \cos{(\omega^0_h \tau_{12})}) \cos{(2p_2 -2\alpha_0)} - \sin{(\omega^0_e \tau_{12})}\sin{(\omega^0_h \tau_{12})}  \right] \cos{(p_3 + p_1 -2 \alpha_0)} \nonumber \\
&+& \left [ \cos{(\omega^0_e \tau_{12})} +\cos{ (\omega^0_h \tau_{12})} \right] \sin{(2p_2 -2 \alpha_0)} \sin{(p_3 + p_1 -2 \alpha_0)},
\eea
where $p_1$ and $p_2$ are directions of polarization of the first and of the second pulses, $p_3$ is the direction of detection polarization, $\alpha_0 = \alpha_{1}-\varphi$ is the angle between eigen polarization $\alpha_1$ and $\mathbf{B}$, $ \omega^0_e$ and $ \omega^0_h$ are the Larmor frequencies of the individual electron and hole or homogeneous ensemble of electrons and holes. $\tau_{12}$ is the delay between the first and second pulses, $T_2$ is the optical phase relaxation time. Here we omit a multiplier describing the pulses power dependence. All angles are counted from the direction of magnetic field. For the experimentally used polarization configurations HHH and HVH, where H polarization corresponds to $p_i = 0$ and V polarization corresponds to $p_i = \pi/2$, one obtains
\be
P^{h}_{HHH} \sim \left[ 1 - 2 \sin^2{(2\alpha_0)} \sin^2{(\omega^0_e \tau_{12}/2)}\sin^2{(\omega^0_h \tau_{12}/2)} \right ] \e^{-\frac{2\tau_{12}}{T_2}},
\ee
\bea
P^{h}_{HVH} &\sim&  [ \cos{([\omega^0_e + \omega^0_h]\tau_{12})} \sin^2{(\alpha_0)} + \cos{([\omega^0_e - \omega^0_h]\tau_{12})} \cos^2{(\alpha_0)}  \nonumber \\
&-& 2 \sin^2{(2\alpha_0)}  \sin^2{(\omega_e \tau_{12}/2)}\sin^2{(\omega_h \tau_{12}/2)}   ] \e^{-\frac{2\tau_{12}}{T_2}}.
\eea
Superscript $h$ in Eqs. (3) and (4) indicates that we do not consider any possible fluctuations of $g$ factors and of angles $\alpha_0$ in the ensemble of QDs. 

Taking into account gaussian distributions of electron and hole spin precession frequencies $\omega^0_{e,h}$ (gaussian distribution of $g$ factors) in the ensemble as follows
\be
F_{e,h}(\omega_{e,h}) = \frac{1}{\sqrt{2\pi } \Delta \omega_{e,h}} \e^{-\frac{1}{2}(\frac{\omega^0_{e,h}-{\omega}_{e,h}}{ \Delta \omega_{e,h}})^2}
\ee
 with the mean values $\hbar {\omega}_{e,h}= \mu_B \tilde{g}_{e,h} B$, standard deviations $\Delta \omega_{e,h} = \mu_B \Delta g_{e,h} B$ ($\Delta g_{e,h}$ is a spread of electron and hole $g$ factor values), and introducing 
 \be
f_e = \e^{-\frac{(\Delta \omega_e \tau_{12})^2}{2}}, f_h = \e^{-\frac{(\Delta \omega_h \tau_{12})^2}{2}}
 \ee
one obtains:
\be
P^{g}_{HHH}  \e^{\frac{2\tau_{12}}{T_2}} \sim  1 -  \frac{1}{2} \sin^2{(2\alpha_0)} [1-  f_e \cos{({\omega}_e \tau_{12})} ] [1- f_h\cos{(\omega_h \tau_{12})} ],
\ee
\bea
P^{g}_{HVH}  \e^{\frac{2\tau_{12}}{T_2}} &\sim&   f_e f_h  \left[ \cos{([\omega_e + \omega_h]\tau_{12})} \sin^2{(\alpha_0)} + \cos{([\omega_e - \omega_h]\tau_{12})} \cos^2{(\alpha_0)} \right ] \nonumber \\
&-& \frac{1}{2} \sin^2{(2\alpha_0)}  [1-  f_e \cos{(\omega_e \tau_{12})} ][ 1- f_h\cos{(\omega_h \tau_{12})} ],
\label{Eq8}
\eea
\bea
\e^{\frac{2\tau_{12}}{T_2}}P^{g} &\sim&  \left[ \sin{(\omega_e \tau_{12})}\sin{(\omega_h \tau_{12})} f_e f_h \cos{(2 p_2-2\alpha_0)} 
+ 1-\cos{(\omega_e \tau_{12})} \cos{(\omega_h \tau_{12})} f_e f_h  \right] \cos{(p_3-p_1)}  \nonumber \\
&+&  [ (1+\cos{(\omega_e \tau_{12})} \cos{(\omega_h \tau_{12})}  f_e f_h) \cos{(2p_2 -2\alpha_0)} \nonumber \\
&-& \sin{(\omega_e \tau_{12})}\sin{(\omega_h \tau_{12})} f_e f_h  ] \cos{(p_3 + p_1 -2 \alpha_0)} \nonumber \\
&+& \left [ \cos{(\omega_e \tau_{12})} f_e +\cos{ (\omega_h \tau_{12})} f_h  \right] \sin{(2p_2 -2 \alpha_0)} \sin{(p_3 + p_1 -2 \alpha_0)},
\label{FitEqs}
\eea
where the superscript $g$ indicates the inhomogeneous broadening of $g$ factor values without a fluctuation of $\alpha_0$ taken into account.

If one assumes that the hole spin is dephased ($f_h=0$) and dephasing of the electron spin is ignored ($f_e=1$), then Eq.~(\ref{Eq8}) can be simplified to Eq. (8) of the main text:
\begin{equation}
P_{HVH} \sim \left (1-\cos{[4(\alpha_1 - \varphi)]} \right) \sin^2{(\omega_e \tau_{12}/2)}.
\label{PHVHinh}
\end{equation}

\section{Data evaluation}

The magnetic field and angular dependences of the PE amplitude at fixed value $\tau_{12}=400$~ps measured in the polarization configuration HHH and HVH are shown in Figs.~\ref{FigColorPlot}(a) and (b). These figures show the angular oscillation in addition to the magnetic field dependent oscillations induced by electron and hole spin precession in transverse magnetic field. Both types of oscillations are discussed in the main text. 

\begin{figure*}
\includegraphics[width=0.48\linewidth]{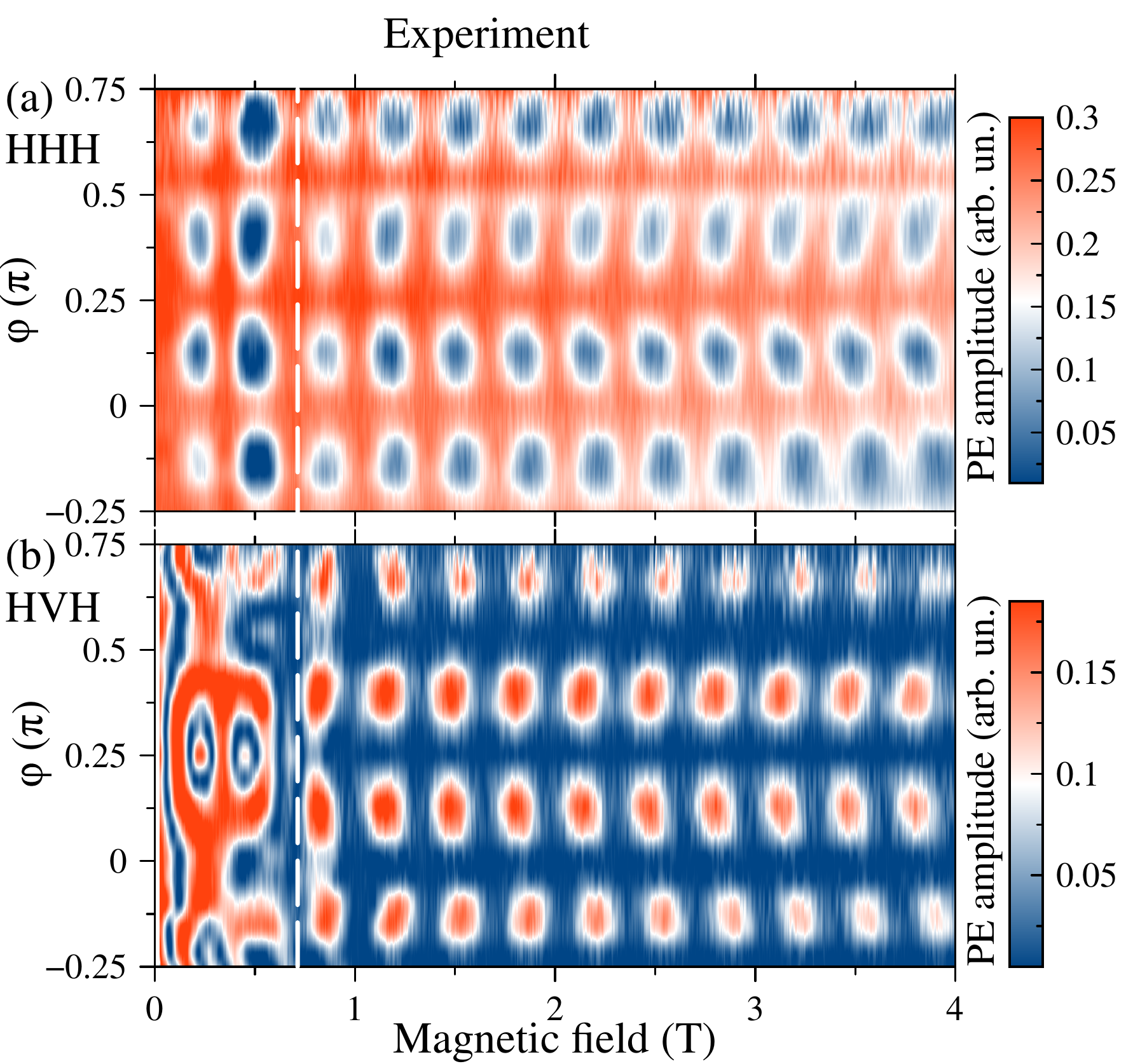}
\includegraphics[width=0.48\linewidth]{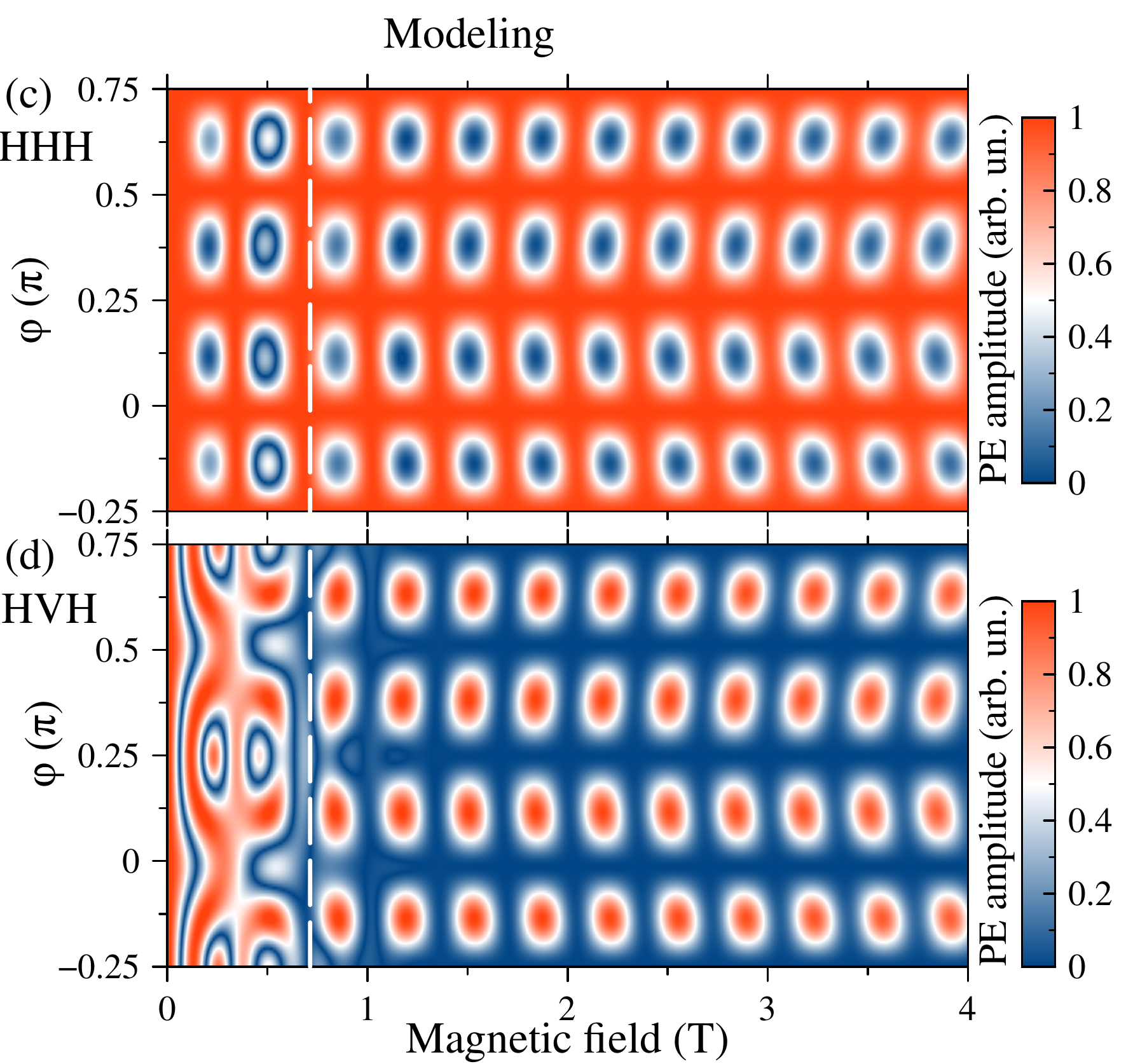}
\caption{Spin-dependent photon echo as function of the magnetic field strength $B$ and sample orientation angle $\varphi$. Experimental figures are measured at HHH (a) and HVH (b) polarization configurations  at $\tau_{12} = 400$~ps. The theoretical figures for HHH (c) and HVH (d) polarization configurations are calculated using $\tau_{12} = 400$~ps, $T_2 = 420$~ps, and parameters for electron and hole $g$-tensor corresponding to the solid lines in Figs.~3 (b) and (c) of the main text, $\Delta g_h = 0.065$, $\Delta g_e = 0.005$ $T_2 = 430$~ps.}
\label{FigColorPlot}
\end{figure*}

For evaluation of the angular dependence of electron and hole $g$ factors shown in Fig.~3(b,c) we used the series of experimental $\tau_{12}$ dependences of PE amplitude  in HHH and HVH polarization configurations measured at $B=0, 0.19, 0.25, 0.33, 0.5, 1, 2, 4$~T at angles $\varphi = -\pi/4, -\pi/8, 0, \pi/8, \pi/4$. We used this set of $B$ keeping in mind the inhomogeneous broadening of hole $g$ factors. The used set of $\varphi$ corresponds to the most informative angles shown in Fig.~\ref{FigColorPlot}(a,b). 

All experimental $\tau_{12}$ dependences are shown in Figures~\ref{0HVH}-\ref{90HHH} by blue lines while corresponding fitting curves by Eq.~\ref{FitEqs} are shown by red dashed lines. For all these curves, the following joint parameters were used: $T_2 = 430$~ps, $\Delta g_h = 0.065$, $\Delta g_e = 0.005$, $p_1 = 0$ (H), $p_2 = 0$ for H and $p_2 = \pi/2$ for V, $p_3 = 0$ (H), $\alpha_0 = -2(\varphi + \delta \varphi) $. For the fittings we assumed small experimental inaccuracy in the orientation of the sample relative to the direction $\bm B$ described by the angle $\delta \varphi$. $\delta \varphi$ also allows us to take into account the influence of the weak low symmetry contribution on $\alpha_0 =\alpha_1 -\varphi$. The variable parameters for series with fixed sample orientation ($\varphi$) and polarization configuration were $\tilde{g}_e$, $\tilde{g}_h$ and $\delta \varphi$ (shown in caption of each figure), but  these parameters were the same for all $B$ values in the series.

Figure~\ref{freqSM}(a,b) shows the obtained angular ($\varphi$) dependence of $\tilde{g}_e$ and $\tilde{g}_h$ by black dots (the same as shown in Fig.~(3) in the main text). Black solid curves are fits of experimental dependencies by Eq.~({4}) in the main text. This allowed us to determine the parameters of electron $g$ factor  $g_e = -0.5312$, $g'_e = 0.0065$, $\chi_e = 0$. Here we take into account the known in literature negative sign of the electron $g$ factor~[27,36,37].
 The angular dependence of the hole $g$ factor corresponds to the values of the parameters $g_h = 0.213$, $g'_h = 0.029$ and $\chi_h =\pi/2$. For these parameters, the maximum deviation of the effective direction of the magnetic field for an electron $\theta_e + \varphi \approx 0.015$ rad and can be neglected indeed. The analogous deviation for the hole is $\theta_h + \varphi \approx 0.15$ rad for $\varphi=0,\pi$ and $\theta_h + \varphi \approx 0$ for $\varphi=\pm \pi/4, \pm 3\pi/4.$

\begin{figure}[h]
\includegraphics[width=\linewidth]{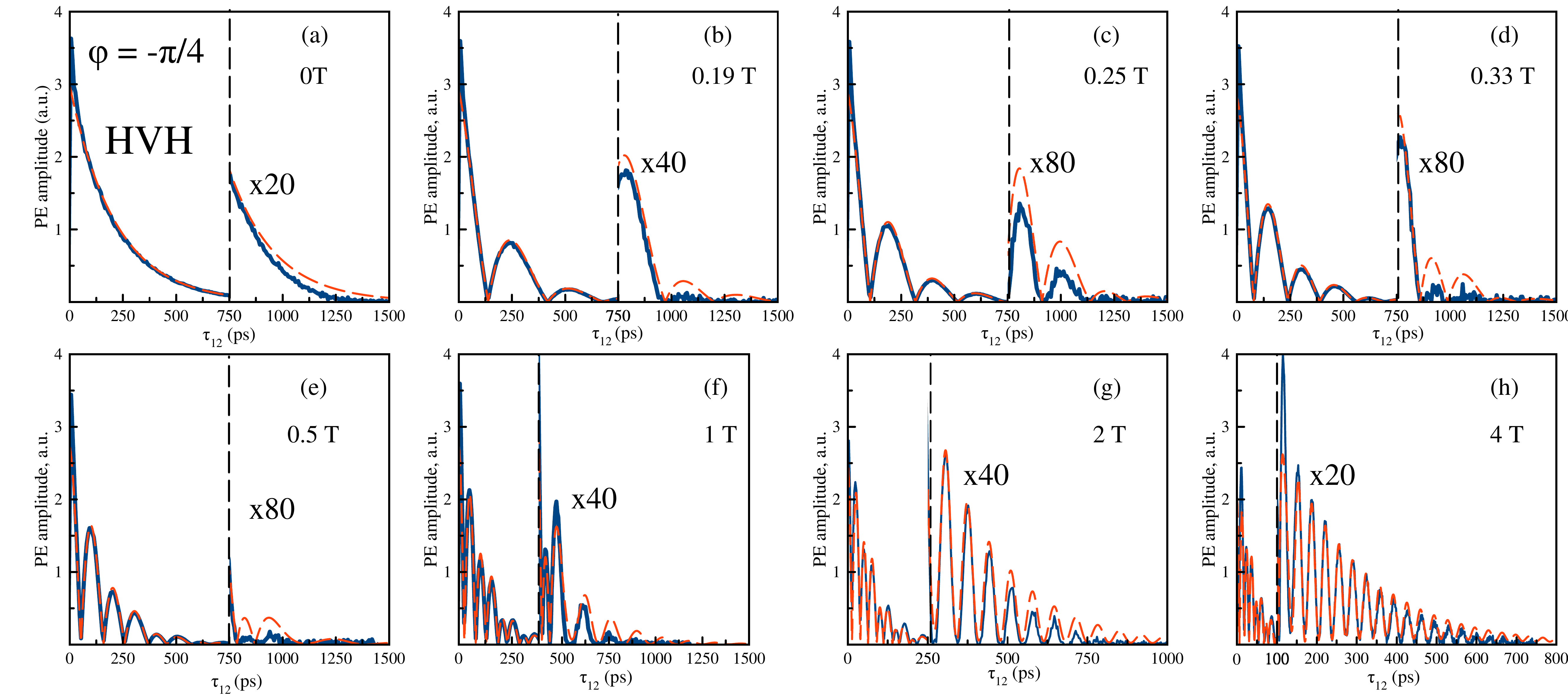}
\caption{(a-h) $\tau_{12}$ dependencies of PE amplitude at $\varphi = -\pi/4$ (blue lines)  in HVH polarization configuration measured at different $B$ and corresponding fitting curves (red dashed lines), $\delta \varphi = 5^\circ$, $\tilde{g}_e = 0.525$, $\tilde{g}_h = 0.181$.
}
\label{0HVH}
\end{figure}

\begin{figure}[h]
\includegraphics[width=\linewidth]{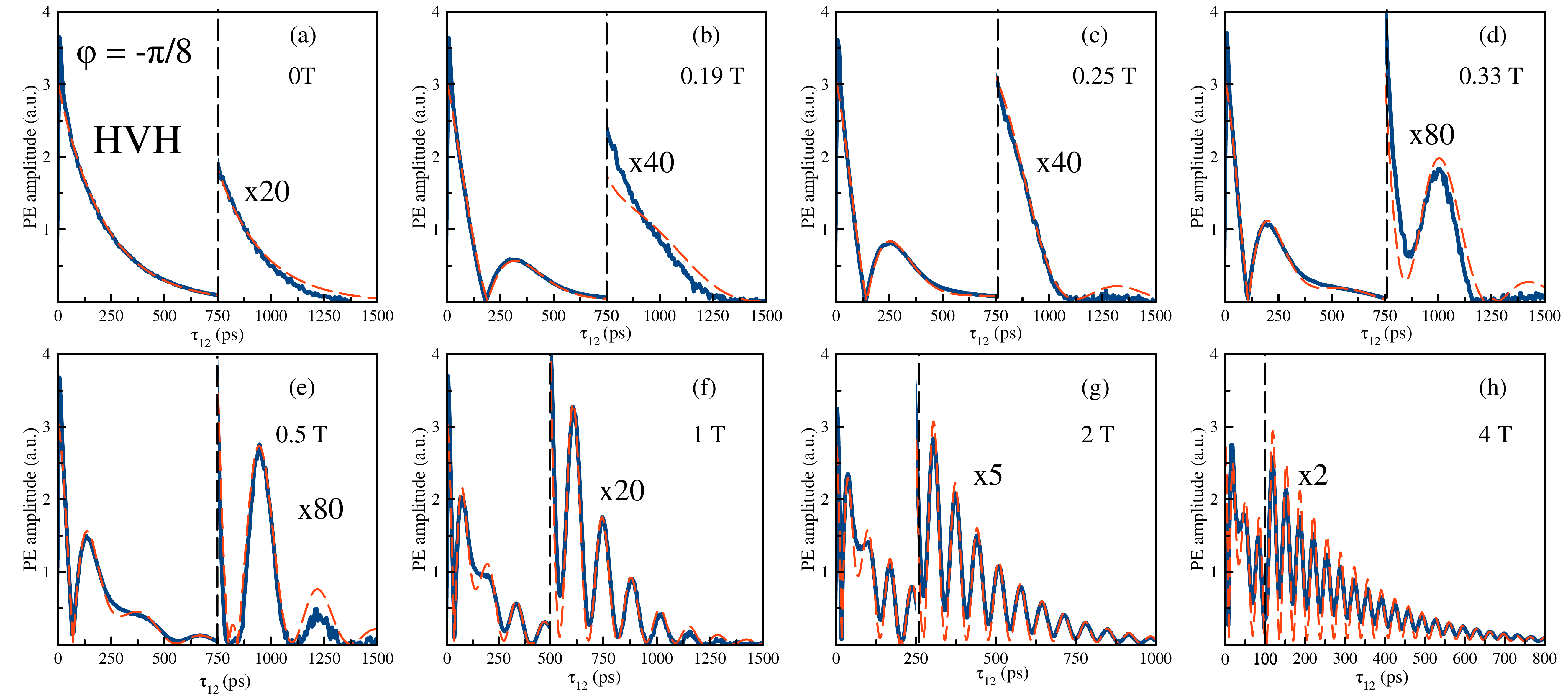}
\caption{(a-h) $\tau_{12}$ dependencies of PE amplitude at $\varphi = -\pi/8$ (blue lines)  in HVH polarization configuration measured at different $B$ and corresponding fitting curves (red dashed lines), $\delta \varphi = 3.5^\circ$, $\tilde{g}_e = 0.526$, $\tilde{g}_h = 0.189$.
}
\label{22HVH}
\end{figure}

\begin{figure}[h]
\includegraphics[width=\linewidth]{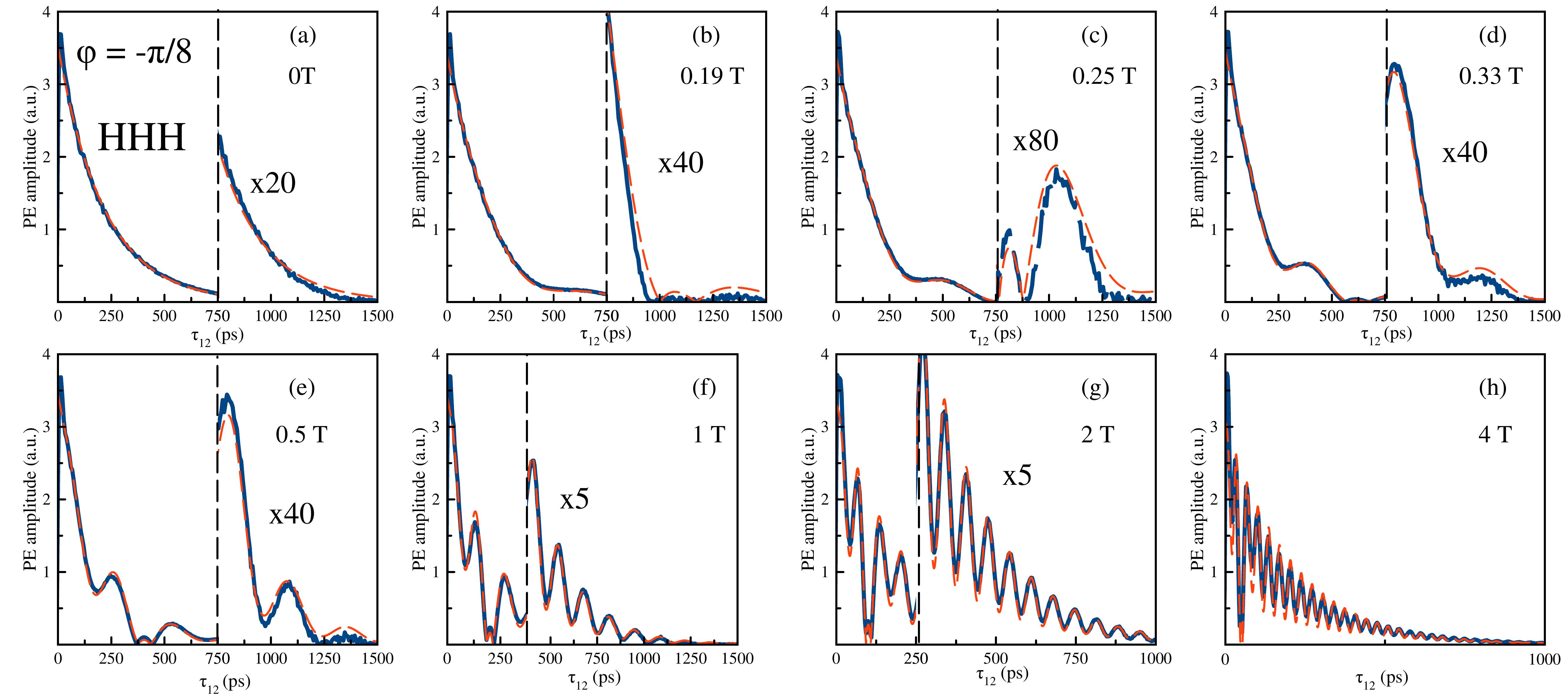}
\caption{(a-h) $\tau_{12}$ dependencies of PE amplitude at $\varphi = -\pi/8$ (blue lines)  in HHH polarization configuration measured at different $B$ and corresponding fitting curves (red dashed lines), $\delta \varphi = 7.5^\circ$, $\tilde{g}_e = 0.526$, $\tilde{g}_h = 0.19$.
}
\label{22HHH}
\end{figure}

\begin{figure}[h]
\includegraphics[width=\linewidth]{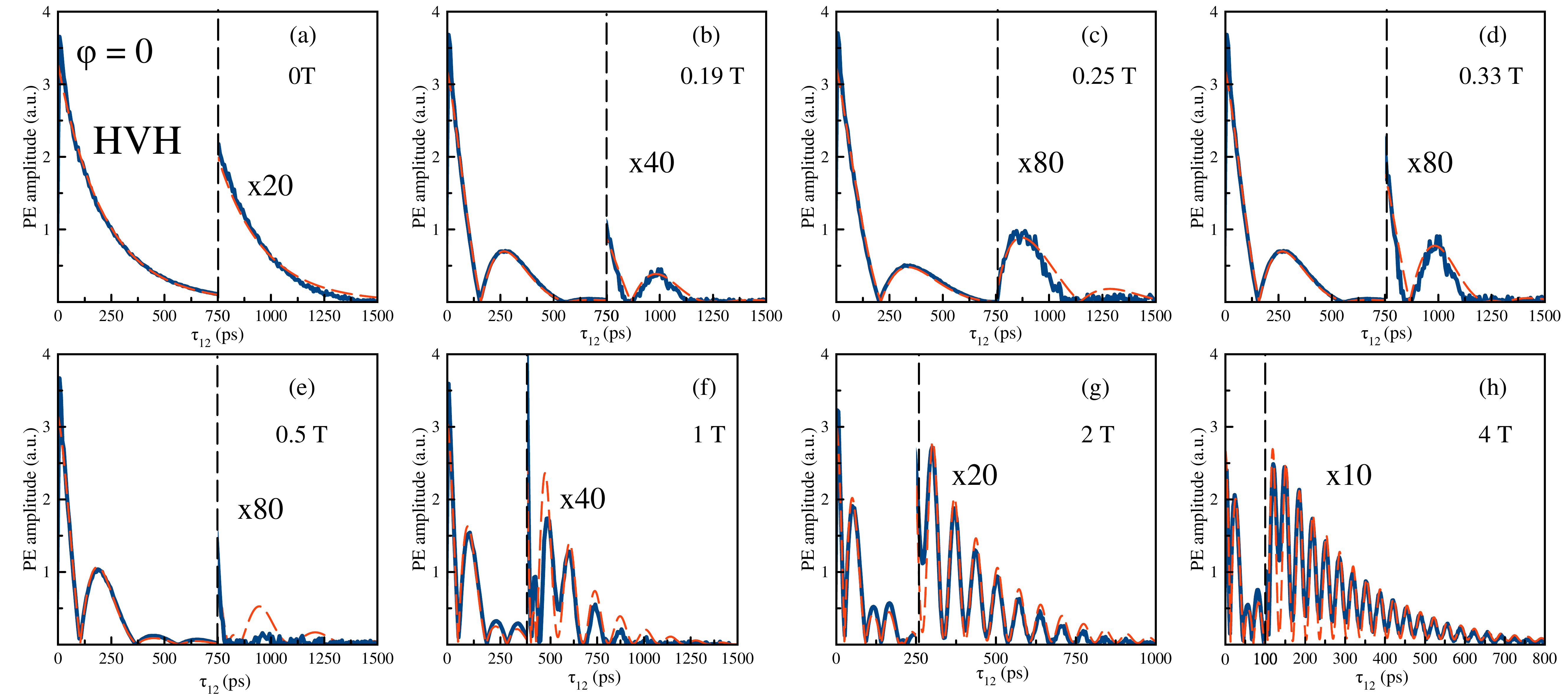}
\caption{(a-h) $\tau_{12}$ dependencies of PE amplitude at $\varphi = 0$ (blue lines)  in HVH polarization configuration measured at different $B$ and corresponding fitting curves (red dashed lines), $\delta \varphi = 7.5^\circ$, $\tilde{g}_e = 0.525$, $\tilde{g}_h = 0.212$.
}
\label{45HVH}
\end{figure}

\begin{figure}[h]
\includegraphics[width=\linewidth]{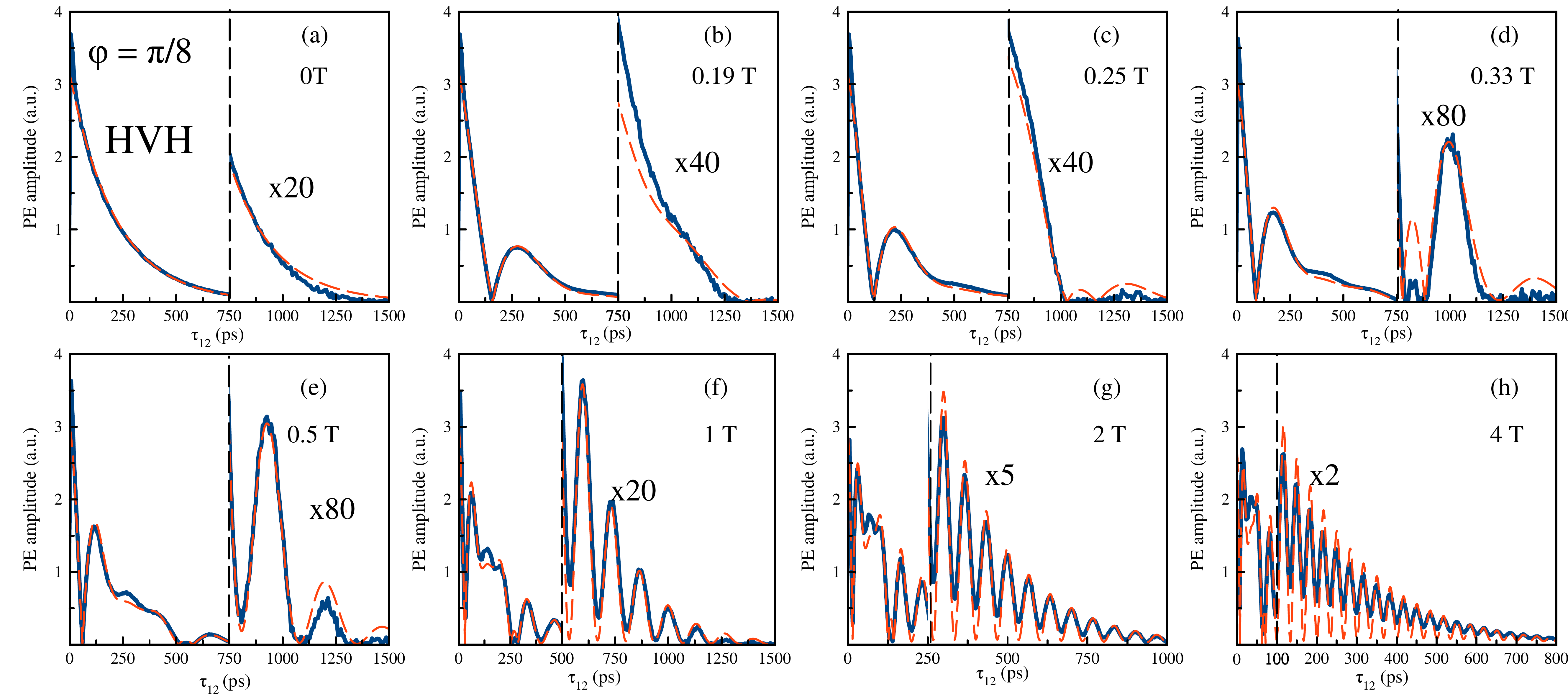}
\caption{(a-h) $\tau_{12}$ dependencies of PE amplitude at $\varphi = \pi/8$ (blue lines)  in HVH polarization configuration measured at different $B$ and corresponding fitting curves (red dashed lines), $\delta \varphi = 2^\circ$, $\tilde{g}_e = 0.535$, $\tilde{g}_h = 0.234$.
}
\label{66HVH}
\end{figure}

\begin{figure}[h]
\includegraphics[width=\linewidth]{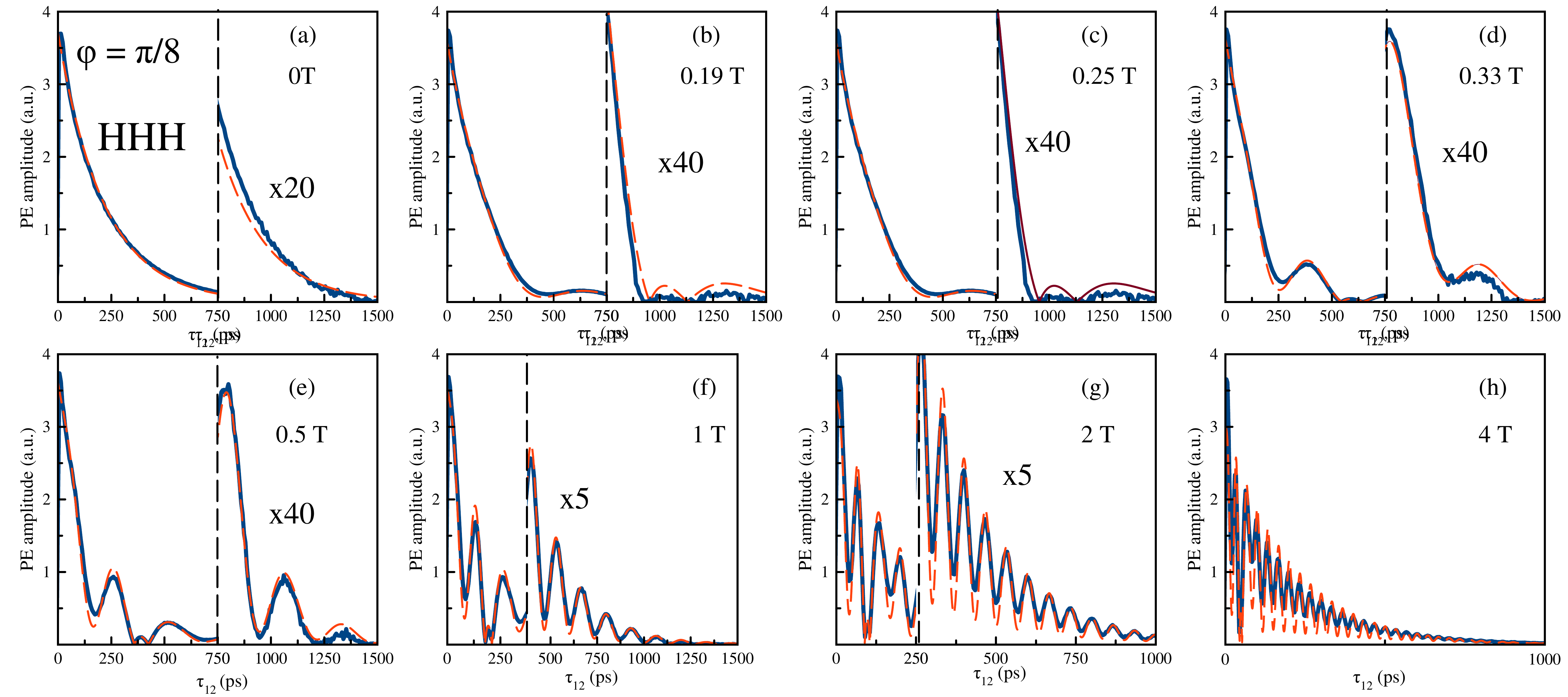}
\caption{(a-h) $\tau_{12}$ dependencies of PE amplitude at $\varphi = \pi/8$ (blue lines)  in HHH polarization configuration measured at different $B$ and corresponding fitting curves (red dashed lines), $\delta \varphi = 5^\circ$, $\tilde{g}_e = 0.535$, $\tilde{g}_h = 0.234$.
}
\label{66HHH}
\end{figure}

\begin{figure}[h]
\includegraphics[width=\linewidth]{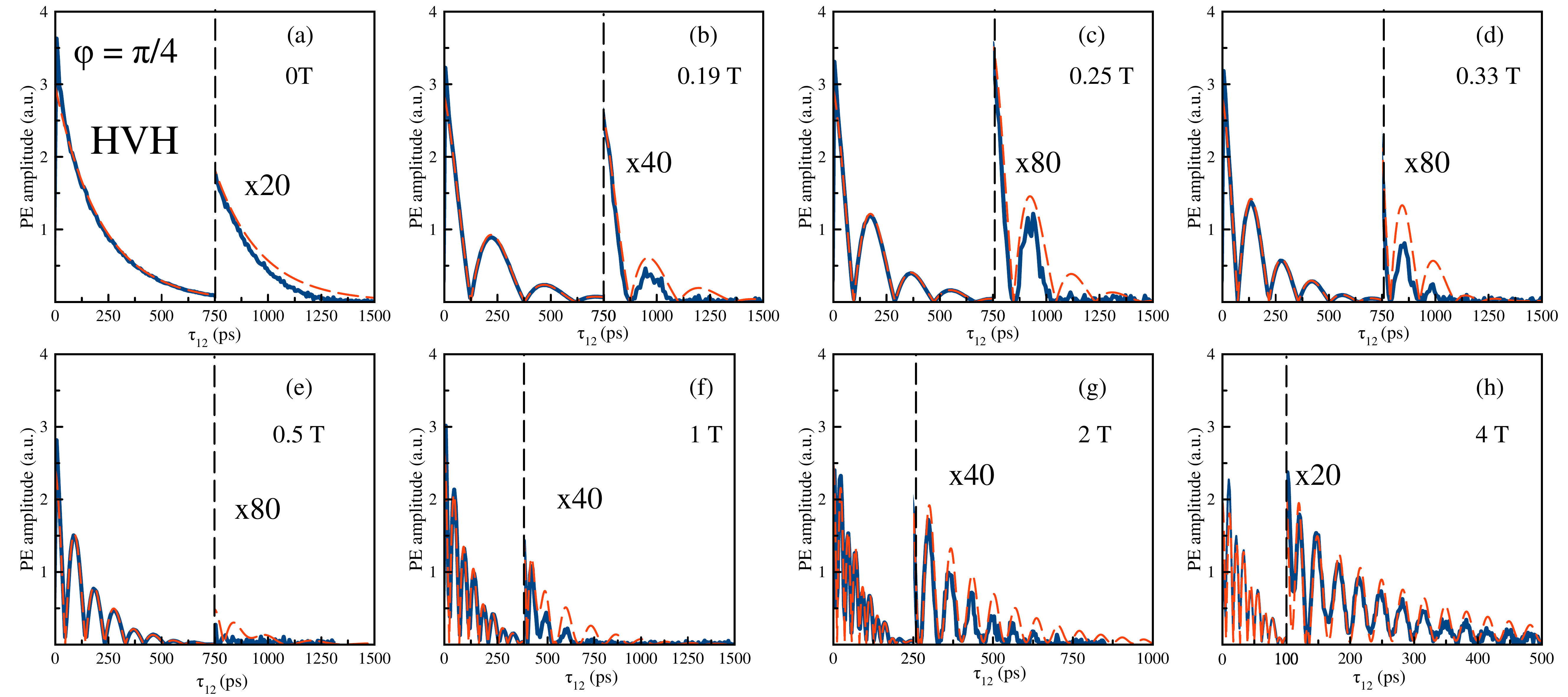}
\caption{(a-h) $\tau_{12}$ dependencies of PE amplitude at $\varphi = \pi/4$ (blue lines)  in HVH polarization configuration measured at different $B$ and corresponding fitting curves (red dashed lines), $\delta \varphi = 4^\circ$, $\tilde{g}_e = 0.538$, $\tilde{g}_h = 0.242$.
}
\label{90HVH}
\end{figure}

\begin{figure}[h]
\includegraphics[width=\linewidth]{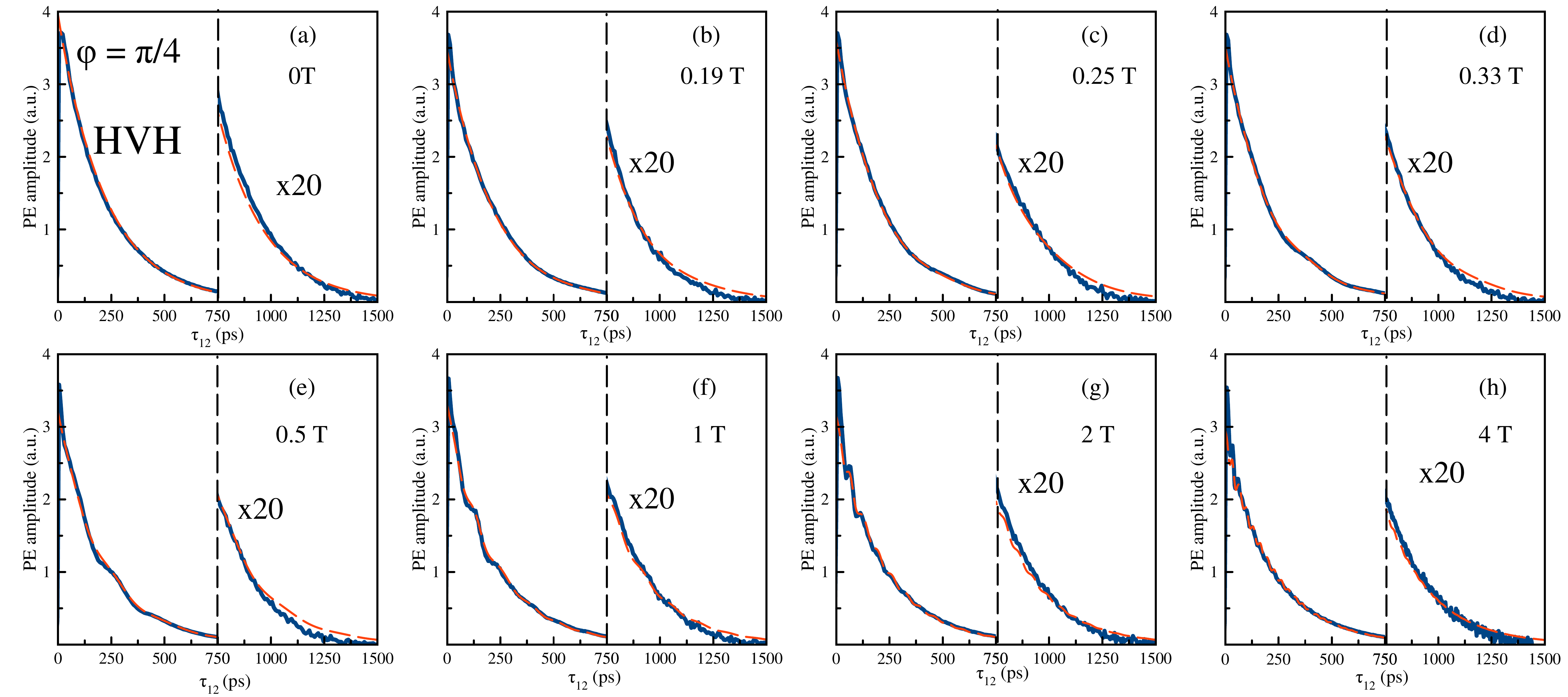}
\caption{(a-h) $\tau_{12}$ dependencies of PE amplitude at $\varphi = \pi/4$ (blue lines)  in HHH polarization configuration measured at different $B$ and corresponding fitting curves (red dashed lines), $\delta \varphi = 4^\circ$, $\tilde{g}_e = 0.538$, $\tilde{g}_h = 0.242$.
}
\label{90HHH}
\end{figure}

\begin{figure}[h]
\includegraphics[width=0.4\linewidth]{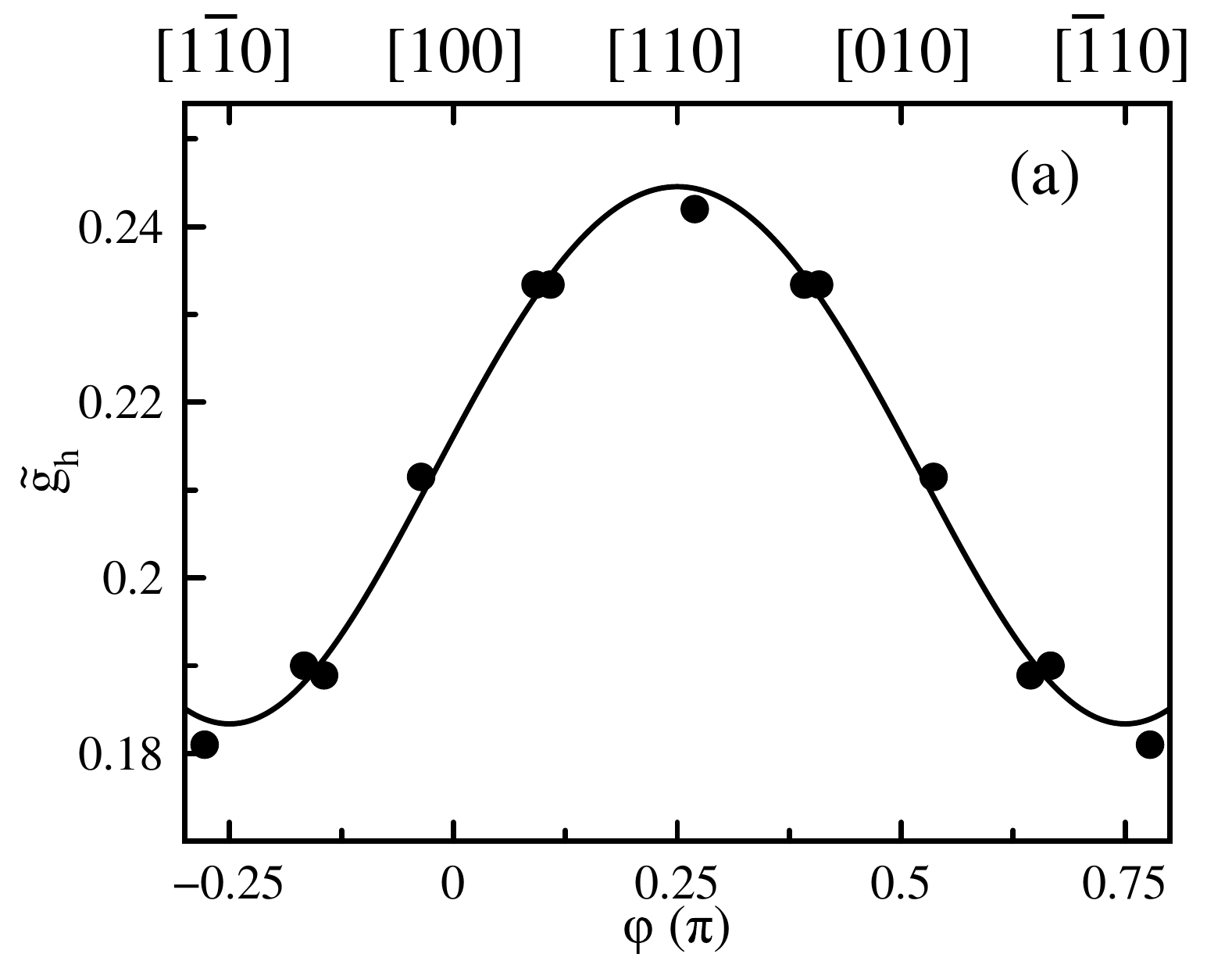}
\includegraphics[width=0.4\linewidth]{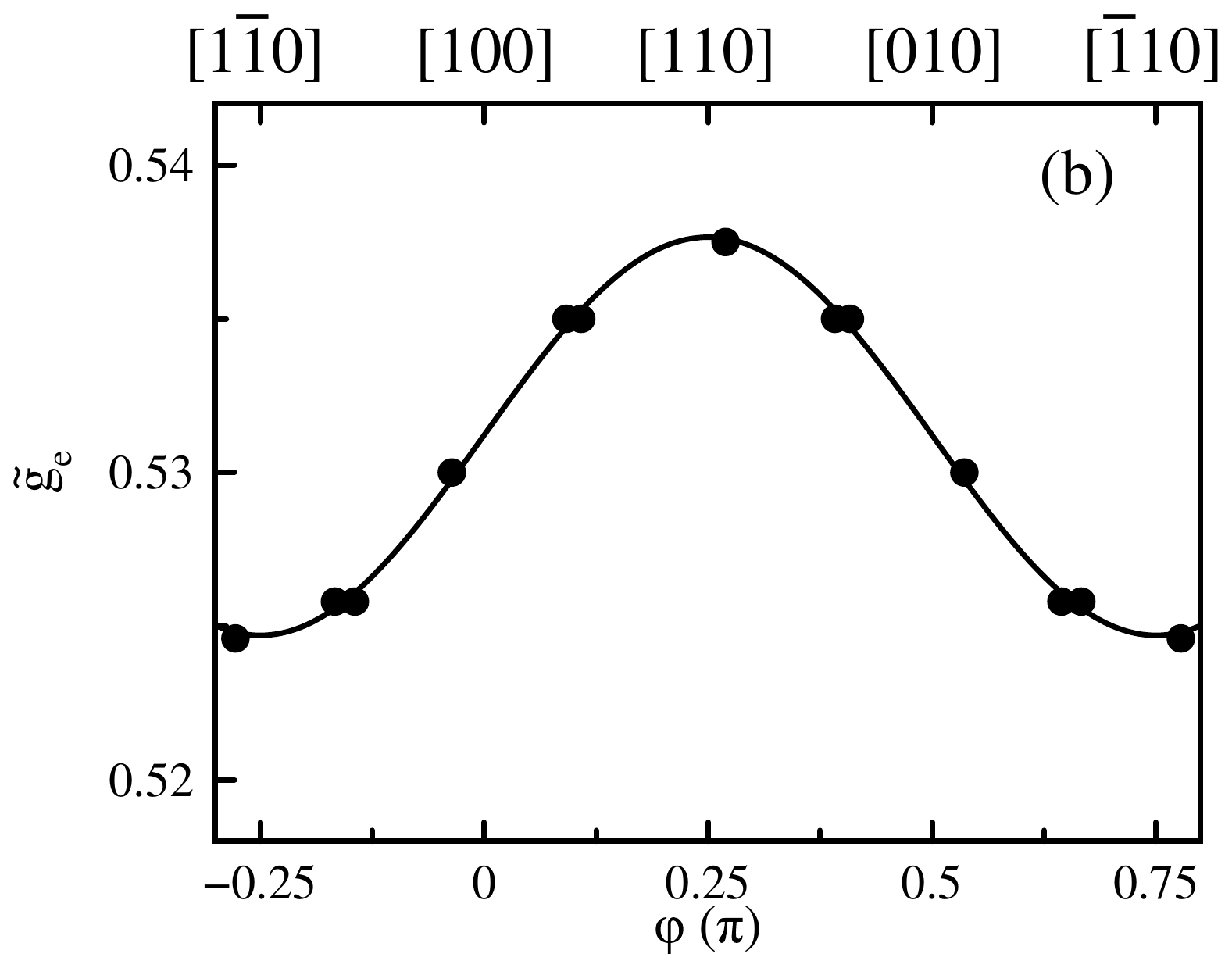}
\caption{ Black dots are angular ($\varphi$) dependence of $\tilde{g}_h$ (a)  and $\tilde{g}_e$ (b) evaluated from Figs.~\ref{0HVH}-\ref{90HVH}. Black solid curves are fit of experimental dependencies by Eq.~(4) in the main text.
}
\label{freqSM}
\end{figure}
 
 \clearpage

\section{In-plane hole $g$-factor in quantum dots}

The in-plane heavy hole $g$-factor in quantum wells caused by their $D_{2d}$ symmetry has the following contribution~[15]:
\begin{equation}
g_{h}^{QW}=  (\gamma_3-\gamma_2)(k_x^2+k_y^2)\tilde{G},
\end{equation}
where
\begin{equation}
\label{tilde_G}
\tilde{G} = 12{\hbar^2 \gamma_3\over m_0}\sum_{n=1}^\infty {z_{hh_1,lh_{2n}}ik^z_{lh_{2n},hh_1}\over E_{hh_1}-E_{lh_{2n}}} .
\end{equation}
Here $\gamma_{2,3}$ are the second and third dimensionless Luttinger parameters, and $k_{x,y}$ are the in-plane components of the hole wavevector. 

\subsection{Rectangular QW  with infinitely high barriers}

For a rectangular QW with infinitely high barriers and width $L_W$ we have
\begin{equation}
z_{1,2n} = {2L_W^2\over \pi^2}{ik^z_{1,2n}\over 4n^2-1}, 
\qquad
k^z_{1,2n} = {4i\over L}{2n\over 1-(2n)^2},
\qquad
z_{1,2n}ik^z_{2n,1} = - {2\over \pi^2}{(8n)^2\over (4n^2-1)^3},
\end{equation}
and
\begin{equation}
E_{hh_1}= {\hbar^2 \over 2m_h} {\pi^2\over L_W^2},
\qquad E_{lh_{2n}} = {\hbar^2 \over 2m_l} {\pi^2(2n)^2\over L_W^2}.
\end{equation}
Here the light- and heavy-hole effective masses for motion along the growth axis [001] are related with the Luttinger valence-band parameters by
\begin{equation}
m_{l,h}= \frac{m_0}{\gamma_1 \pm 2 \gamma_2}\:,
\end{equation}
where $m_0$ is the free electron mass.

As a result, the value $\tilde{G}$ can be presented as $\tilde{G} = G L_W^2$ where $G$ is given by  Eq.~(9) of Ref.~[15]:
\bea
G={1024\over \pi^4} \sum_{n=1}^\infty {3n^2\gamma_3\over (4n^2-1)^3[4(\gamma_1+2\gamma_2)n^2-\gamma_1+2\gamma_2]} \\
 ={12\gamma_3 \over \pi^2(\gamma_1+2\gamma_2)} \left[ {16\sqrt{\beta}\over \pi}\ctg{\left({\pi\over 2}\sqrt{\beta}\right)} -3 + 2\beta + \beta^2\right] \nonumber.
\eea
Here we introduced the effective mass ratio 
\begin{equation}
\label{beta}
\beta = {m_l\over m_h}
={\gamma_1 - 2\gamma_2 \over \gamma_1 + 2 \gamma_2}\:.
\end{equation}

For both GaAs ($\gamma_1=6.8$, $\gamma_2=2.1$, $\gamma_3=2.9$) and InAs ($\gamma_1=19.7$, $\gamma_2=8.4$, $\gamma_3=9.3$) we have $(\gamma_3-\gamma_2)G\approx 0.07$, and
\begin{equation}
g_h^{QW} \approx 0.08(k_x^2+k_y^2)L_W^2.
\end{equation}

In QDs, the in-plane motion is quantized.  In the model of a cylindrical QD with a disk radius $R$, we obtain another expression:
\begin{equation}
\left<k_x^2 + k_y^2 \right> = {\alpha^2\over R^2}.
\end{equation}
Here $\alpha$ is the first root of the zero-order Bessel function: $J_0(\alpha)=0$, $\alpha \approx 2.405$.
In this model we have for the heavy-hole in-plane $g$-factor in a QD
\begin{equation}
g_h^{QD} \approx 0.4{L_W^2\over R^2}.
\end{equation}

\subsection{Parabolic QW}

Let us now consider a QW model with a parabolic potential
\begin{equation}
U(z) = {\kappa_z z^2\over 2}.
\end{equation}
In this case {the set of quantum confined eigen functions for a particle with the effective mass $m$ are}
\begin{equation}
\psi_N(z)={1\over \pi^{1/4}\sqrt{L} \sqrt{2^N N!}}\text{e}^{-z^2/2L^2} H_N(z/L),
\qquad L= \sqrt[4]{ \frac{\hbar^2}{\kappa_z m} }.
\end{equation}
Here $N=0,1,2\ldots$, and energies $E_N=\hbar \sqrt{\kappa_z\over m}(N+1/2)$.
It is convenient to introduce the index $\nu=N+1$ where $\nu=1,2,3\ldots$ In terms of $\nu$, ${\psi_\nu(z)=\psi_{N+1}(z)}$, and the eigen energies are
$$E_\nu=\hbar \sqrt{\kappa_z\over m}(\nu-1/2)\:.$$ 

The wavefunctions of the ground heavy-hole subband $hh_1$ and even light-hole subbands $lh_{2n}$
($n=1,2,3\ldots$) have the form 
\begin{equation}
\psi_{hh_1}={1\over \pi^{1/4}\sqrt{L_h}}\text{e}^{-z^2/2L_h^2}\:,
\qquad
\psi_{lh_{2n}}={1\over \pi^{1/4}\sqrt{L_l} \sqrt{2^{2n-1} (2n-1)!}}\text{e}^{-z^2/2L_l^2} H_{2n-1}(z/L_l)\:,
\end{equation}
where $L_l$ and $L_h$ are the light- and heavy-hole lengths
\begin{equation}
L_{l,h}= \sqrt[4]{ \frac{\hbar^2}{\kappa_z m_{l,h}} }\:.
\end{equation}

The calculation yields
\begin{equation}
z_{hh_1,lh_{2n}}={\tilde{L}^3\over L_l\sqrt{L_lL_h}} {\sqrt{(2n-1)!} \over \sqrt{2^{2n-1}}(n-1)!}
	\left[ \left({\tilde{L}\over L_l}\right)^2-1\right]^{n-1},  \qquad ik^z_{lh_{2n},hh_1} = \frac{ z_{hh_1, lh_{2n}} }{L_h^2},
\end{equation}
where the length $\tilde{L}$ is defined by
\begin{equation}
{1\over\tilde{L}^2} = {1\over 2} \left( {1\over L_l^2}+{1\over L_h^2}\right).
\end{equation}
Therefore Eq.~\eqref{tilde_G} yields
\begin{equation}
\tilde{G} = 24{\hbar \gamma_3\over m_0}{\tilde{L}^6\over L_l^3 L_h^3}  \sqrt{m_l\over \kappa_z}\sum_{n=1}^\infty 
{(2n-1)! \over (4n-1 -\sqrt{m_l/ m_h}) 2^{2n-1}[(n-1)!]^2}
\left[ \left({\tilde{L}\over L_l}\right)^2-1\right]^{2n-2}\:.
\end{equation}
Then we obtain
\begin{equation}
\tilde{G} = GL_0^2,
\qquad
\left({\tilde{L}\over L_l}\right)^2 = {2\sqrt{\beta}\over 1+\sqrt{\beta}},\quad
\left({\tilde{L}\over L_h}\right)^2 = {2 \over 1+\sqrt{\beta}},\quad
{\tilde{L}^2\over L_lL_h}= {4\sqrt{\beta}\over (1+\sqrt{\beta})^2},
\end{equation}
where the effective mass ratio $\beta$ is given by Eq.~\eqref{beta}, and
the free-electron characteristic length is
\begin{equation}
L_0= \sqrt[4]{ \frac{\hbar^2}{\kappa m_0} }\:.
\end{equation}

Finally, using the relations
\begin{equation}
{(2n-1)! \over 2^{2n-1}[(n-1)!]^2} = {\Gamma(2n) \over 2^{2n-1}[\Gamma(n)]^2} ={1\over \sqrt{\pi}} {\Gamma(n+1/2)\over\Gamma(n)},
\end{equation}
and
\begin{equation}
\sum_{n=1}^\infty {\zeta^{n}\over  n-x}
{\Gamma(n+1/2)\over\Gamma(n)} = {\sqrt{\pi}\zeta\over 2(1-x)}\: _2F_1\left({3\over 2},1-x;2-x;\zeta \right),
\end{equation}
where $_2F_1(\alpha,\beta;\gamma;z)$ is the hypergeometric function,
we get
\begin{equation}
G ={3\gamma_3 2^8 \over \sqrt{\gamma_1+2\gamma_2} (3-\sqrt{\beta})} {\beta^{3/2}\over (1+\sqrt{\beta})^6}
  \: _2F_1\left({3\over 2},{3-\sqrt{\beta}\over 4};{7-\sqrt{\beta}\over 4};\left( {1-\sqrt{\beta}\over 1+\sqrt{\beta}}\right)^2 \right).
\end{equation}

If we assume the in-plane QD potential to be also parabolic~[40]:
\begin{equation}
U(\rho) = {\kappa_\parallel \rho^2\over 2},
\end{equation}
then the ground in-plane state has the following wavefunction and energy
\begin{equation}
\psi(\rho) = {\exp(-\rho^2/L_\parallel^2)\over \sqrt{\pi}L_\parallel}, \qquad E=\hbar\sqrt{\kappa_\parallel\over m_\parallel}, \qquad L_\parallel= \sqrt[4]{ \frac{\hbar^2}{\kappa_{\parallel} m_{\parallel}} },
\end{equation}
where $m_\parallel$ is the in-plane effective mass. 
Averaging the sum $k_x^2 + k_y^2$ over this state yields
\begin{equation}
\left< k_x^2+k_y^2\right> = {1\over L_\parallel^2}= {1\over L_0^2}\sqrt{ \kappa_\parallel m_\parallel\over \kappa m_0}.
\end{equation}
For the heavy hole $m_\parallel=m_0/(\gamma_1+\gamma_2)$, therefore we have
\begin{align}
&g_{h}^{QD}= \sqrt{\kappa_\parallel \over \kappa_z} G_{\rm parab},
\\ 
&G_{\rm parab}= {\gamma_3-\gamma_2  \over \sqrt{ (\gamma_1+\gamma_2)(\gamma_1+2\gamma_2)}} {3\gamma_3 2^8 \over (3-\sqrt{\beta})} {\beta^{3/2}\over (1+\sqrt{\beta})^6}
  \: _2F_1\left({3\over 2},{3-\sqrt{\beta}\over 4};{7-\sqrt{\beta}\over 4};\left( {1-\sqrt{\beta}\over 1+\sqrt{\beta}}\right)^2 \right). \nonumber
\end{align}

For the parameters of GaAs we obtain the value $G_{\rm parab} \approx 0.5$.

\end{widetext}

\end{document}